\documentclass[a4paper,fleqn]{cas-sc}
\usepackage{amsmath,amsfonts}
\usepackage{algorithmic}
\usepackage{algorithm}
\usepackage{array}
\usepackage{textcomp}
\usepackage{stfloats}
\usepackage{url}
\usepackage{verbatim}
\usepackage{graphicx}
\usepackage{cite}
\usepackage{caption}
\usepackage{amsthm}
\usepackage{enumitem}
\usepackage{subcaption}
\usepackage{siunitx}
\usepackage{booktabs}
\usepackage{tikz}
\usepackage{pgfplots} 
\usepackage{longtable}
\pgfplotsset{compat=1.16} 
\usepackage{balance}
\usepackage{bigstrut,array,multirow,tabularx}
\usepackage{pifont}
\usepackage{dsfont}
\usepackage{mathtools}
\usepackage{makecell} 
\usepackage{tabularray}
\pgfplotsset{compat=newest}
\usepackage{anyfontsize}
\usepackage{setspace}
\usepackage{amsthm}
\usepgfplotslibrary{units}
\usetikzlibrary{patterns, shapes.geometric, positioning}

\usepackage[authoryear]{natbib}
\renewcommand{\cite}{\citep}

\usepackage{array,tabularx}
\newcolumntype{Y}{>{\centering\arraybackslash}X} 

\newtheorem{theorem}{Theorem}[section]  

\newproof{pf}{Proof}
\newproof{pot}{Proof of Theorem \ref{thm}}

\begin{document}
\let\WriteBookmarks\relax
\def\floatpagepagefraction{1}
\def\textpagefraction{.001}

\title [mode = title]{Training-free Adjustable Polynomial Graph Filtering for Ultra-fast Multimodal Recommendation}  

\shorttitle{Graph Filtering for Multimodal Recommendation}

\shortauthors{Yu-Seung Roh et~al.}

%

\author[1]{Yu-Seung Roh}

\author[2]{Joo-Young Kim}

\author[3]{Jin-Duk Park}

\author[1]{Won-Yong Shin}\cormark[1]

\affiliation[1]{addressline={School of Mathematics and Computing (Computational Science and Engineering)}, 
            organization={Yonsei University},
            city={Seoul},
            postcode={03722},
            country={Republic of Korea}}
\affiliation[2]{addressline={Graduate School of Data Science}, 
            organization={Seoul National University},
            city={Seoul},
            postcode={08826},
            country={Republic of Korea}}
\affiliation[3]{organization={NAVER Corporations},
            city={Bundang},
            postcode={13561},
            country={Republic of Korea}}


\ead{wy.shin@yonsei.ac.kr}

\cortext[1]{Corresponding author}


\begin{abstract}
Multimodal recommender systems improve the performance of canonical recommender systems with no item features by utilizing diverse content types such as text, images, and videos, while alleviating inherent sparsity of user--item interactions and accelerating user engagement. However, current neural network-based models often incur significant computational overhead due to the complex training process required to learn and integrate information from multiple modalities. To address this challenge, we propose a \textit{training-free} multimodal recommendation method grounded in \textit{graph filtering}, designed for multimodal recommendation systems to achieve efficient and accurate recommendation. Specifically, the proposed method first constructs multiple similarity graphs for two distinct modalities as well as user--item interaction data. Then, it optimally fuses these multimodal signals using a polynomial graph filter that allows for precise control of the frequency response by adjusting frequency bounds. Furthermore, the filter coefficients are treated as hyperparameters, enabling flexible and data-driven adaptation. Extensive experiments on real-world benchmark datasets demonstrate that the proposed method not only improves recommendation accuracy by up to 22.25\% compared to the best competitor but also dramatically reduces computational costs by achieving \textbf{\underline{the runtime of less than 10 seconds.}}
\end{abstract}


\maketitle

\section{Introduction}

\subsection{Background and Motivation}
\label{section1.1}
Recently, multimodal recommender systems (MRSs) have garnered significant attention due to their ability to accommodate diverse item information from multiple modalities for enhanced recommendation performance. Compared to canonical recommender systems \cite{ranjbar2015imputation, DBLP:journals/eaai/ValcarceLPB19, DBLP:conf/sigir/0001DWLZ020} with no item features (referred to as single-modal recommender systems), MRSs can capture and leverage precise item information (e.g., textual and/or visual features) for recommendations, thereby enhancing the overall capabilities of recommender systems \cite{DBLP:conf/aaai/HeM16, DBLP:conf/mm/WeiWN0C20}. Notably, while single-modal recommender systems often struggle with sparsity of user--item interactions, MRSs can overcome this limitation by utilizing multimodal features of items. Due to these advantages, MRSs \cite{DBLP:conf/mm/Zhang00WWW21, DBLP:conf/www/ZhouZLZMWYJ23, DBLP:conf/mm/ZhouS23, DBLP:conf/mm/Yu0LB23} are shown to substantially outperform single-modal recommendation methods based on collaborative filtering (CF) that rely solely on historical user--item interactions.

Various MRSs have been developed to improve recommendation performance. In particular, thanks to the expressive capability via message passing in graph convolutional networks (GCNs) \cite{DBLP:conf/iclr/KipfW17}, attention has been paid to GCN-based MRSs \cite{DBLP:conf/mm/WeiWN0C20, DBLP:conf/mm/Zhang00WWW21, DBLP:conf/www/ZhouZLZMWYJ23, DBLP:conf/mm/ZhouS23, DBLP:conf/mm/Yu0LB23}. For example, prior studies on MRSs learned GCNs separately to process different modalities, distinct from user--item interactions \cite{DBLP:conf/mm/WeiWN0C20, DBLP:conf/mm/Zhang00WWW21, DBLP:conf/wsdm/KimKSK24}, or constructed an item--item similarity graph based on multimodal information (i.e., multimodal features of items) and then applied GCNs to process the similarity graph \cite{DBLP:conf/mm/Zhang00WWW21}.

\begin{figure}[t]
    \centering
    \includegraphics[width=0.7\columnwidth]{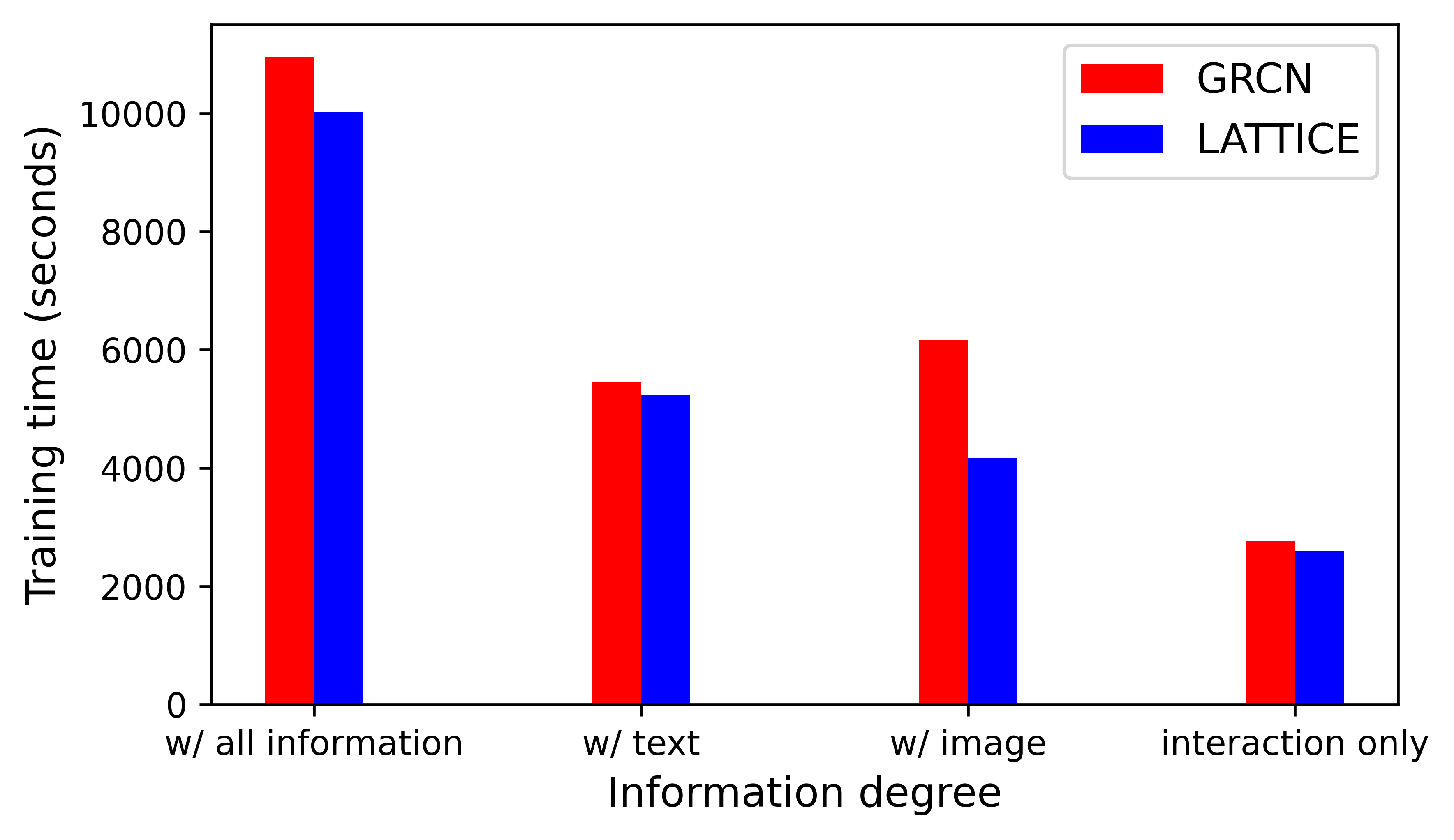}
    \vspace{-1.5mm}
    \caption{Training time comparison of two GCN-based MRSs under different degrees of modality information on the Baby dataset.}
    \vspace{-3mm}
    \label{model_feature}
\end{figure}

On one hand, user preferences tend to shift quickly under the influence of trends, personal situations, and exposure to new content \cite{ju2015using, DBLP:journals/ml/PereiraGAO18}; hence, recommender systems need to be flexible to adapt to such dynamic preferences. Especially in environments where the training and inference speed is crucial, model runtime can become a significant bottleneck. In this context, incorporating additional information (i.e., multimodal features of items) into modeling significantly increases computational overhead for GCN-based MRSs. Consequently, MRSs learn multimodal information through GCNs, naturally leading to escalation of training time for GCNs with the increased number of modalities. Figure \ref{model_feature} illustrates how the training time (in seconds) of various MRSs behaves according to different degrees of modality information on the Baby dataset (one of well-known benchmark datasets for MRSs). As shown in Figure \ref{model_feature}, incorporating all modalities substantially increases the training time of GCN-based MRSs such as GRCN \cite{DBLP:conf/mm/WeiWN0C20} and LATTICE \cite{DBLP:conf/mm/Zhang00WWW21}, highlighting the heavy computational burden of training-based MRSs. Thus, although GCN-based MRSs (e.g., \cite{DBLP:conf/mm/WeiWN0C20}, \cite{DBLP:conf/mm/Zhang00WWW21}) are promising, their computational cost becomes prohibitive in scenarios where models must be updated frequently to track rapidly evolving user preferences.

On the other hand, recent studies on single-modal recommender systems \cite{DBLP:conf/cikm/ShenWZSZLL21, DBLP:conf/www/XiaLGLLG22, DBLP:conf/www/LiuLGL0SG23, DBLP:conf/sigir/ParkSS24} have adopted the {\it training-free} graph filtering (GF) (also known as graph signal processing) mechanism due to its simplicity and effectiveness. As training-free designs, GF bypasses iterative gradient-based training and instead performs graph filtering along with similarity graphs, thereby resulting in substantially lower computational overhead and much faster inference for recommendation than the case of conventional deep or graph neural network (GNN)-based models. Such GF is a fundamental operation enabling the manipulation of signals defined over the nodes based on the underlying graph topology \cite{DBLP:conf/nips/DefferrardBV16, DBLP:journals/tsp/IsufiGSS24}. A pioneering study on training-free GF is GF-CF \cite{DBLP:conf/cikm/ShenWZSZLL21}, which presented a novel approach to constructing an item--item similarity graph and applying GF that does not necessitate model training to enhance recommendation performance. Additionally, recent studies \cite{DBLP:conf/sigir/ParkSS24, DBLP:conf/www/ParkYS25, DBLP:conf/www/KimCPS25} proposed a normalization scheme with tunable user--item influence and smoothing strength, combined with polynomial graph filters that avoid explicit eigendecomposition.

Despite these advances, the GF-based recommendation methods in \cite{DBLP:conf/sigir/ParkSS24, DBLP:conf/www/ParkYS25, DBLP:conf/www/KimCPS25} present a key limitation from the perspective of frequency analysis. Normalization techniques in \cite{DBLP:conf/cikm/ShenWZSZLL21, DBLP:conf/www/XiaLGLLG22, DBLP:conf/www/LiuLGL0SG23} ensure that all eigenvalues of the underlying item--item similarity graph lie within the interval $[0, 1]$, providing a stable theoretical foundation for designing polynomial filters in the frequency domain. In contrast, the normalization and additional adjustment process in \cite{DBLP:conf/sigir/ParkSS24, DBLP:conf/www/ParkYS25, DBLP:conf/www/KimCPS25} does not enforce this spectral constraint, allowing eigenvalues to extend beyond the unit interval, which distorts the intended frequency response. This effect is magnified on multimodal information due to heterogeneous characteristics, resulting in greater spectral instability than the case of user–item interactions. Additionally, such an effect can be amplified with higher-order polynomials, disproportionately emphasizing components tied to large-magnitude eigenvalues and further degrading frequency control. The aggregation of multiple out-of-band components generates spurious, highly distorted signals. These limitations constitute a fundamental theoretical flaw that calls for a principled reformulation of polynomial graph filtering in recommendation. Moreover, their polynomial filters in \cite{DBLP:conf/sigir/ParkSS24, DBLP:conf/www/ParkYS25, DBLP:conf/www/KimCPS25} were analytically derived from a {\it fixed} frequency response function, which lacks flexibility. This limitation underscores the need for a more adjustable and theoretically grounded filtering strategy for recommender systems.

\subsection{Main Contributions}

To tackle the aforementioned challenges, we propose \textsf{MultiModal-Graph Filtering} (\textsf{MM-GF}), a new GF method tailored for MRSs. Specifically, \textsf{MM-GF} constructs multiple similarity graphs derived from user--item interaction data and two distinct modalities. Unlike user--item interactions, multimodal information (i.e., multimodal features of items) often contains {\it outliers} and {\it negative values}, which can make the resulting similarity graphs highly sensitive to noise or prone to singularities (e.g., division by zero during normalization). To mitigate these issues, \textsf{MM-GF} leverages similarity calculations within each modality, thereby capturing modality-specific item--item relationships in a stable manner.

In the GF process, we theoretically identify a critical issue in regard to eigenvalue bounds (see Section \ref{4.1.Motivation} for technical details). \textsf{MM-GF} employs a polynomial graph filter that enables {\it precise frequency control} by adjusting the spectral bounds, thereby eliminating the limitations associated with unbounded eigenvalues. The filter coefficients are treated as tunable hyperparameters, allowing for flexible, data-driven adaptation of the frequency response to match the spectral properties of a given dataset. This design facilitates seamless adaptation to diverse spectral properties of the given data, enhancing the filter’s effectiveness across diverse recommendation scenarios, and ultimately allowing \textsf{MM-GF} to optimally aggregate multimodal information. 

Extensive experiments on various real-world datasets demonstrate that \textsf{MM-GF} achieves up to $22.25$\% higher accuracy and up to \underline{$\times 100.4$ faster runtime} compared to the best MRS competitor. In other words, \textsf{MM-GF} is not only extremely fast but also highly accurate, compared to the GCN-based MRSs harnessing computationally expensive model training.

Our contributions are summarized as follows:
\begin{itemize}
    \item {\bf Theoretical finding}: We theoretically investigate the impact of asymmetric normalization and adjustment on the eigenvalue bounds. This analysis reveals that even minor changes in normalization can significantly shift the spectral range, which in turn affects the stability of GF.
    
    \item {\bf Methodology}: We devise \textsf{MM-GF}, a new and theoretically sound polynomial graph filter that precisely adjusts spectral bounds, ensuring accurate frequency design control and resolving the eigenvalue constraint issue in asymmetric graph constructions with adjustment parameters.

    \item {\bf Extensive evaluation}: We carry out comprehensive experiments, which include cold-start and noisy feature settings, on three widely used benchmark datasets for MRSs. The results consistently demonstrate the effectiveness of \textsf{MM-GF} in terms of computational complexity and model accuracy, compared to GCN-based MRSs.
\end{itemize}

\subsection{Organization and Notations}
The remainder of this paper is organized as follows. In Section \ref{section 2}, we summarize prior work relevant to our study. Section \ref{section 3} provides some preliminaries such as the notion of GF and the problem definition. Section \ref{section 4} describes the technical details of the proposed \textsf{MM-GF} method. Extensive experimental results and analyses are presented in Section \ref{section 5}. Finally, we provide a summary and concluding remarks in Section \ref{section 6}.

\begin{table}[b]
\centering
\footnotesize
  \caption{Summary of notations.}
  \begin{tabular}{cl}
    \toprule
    \bf Notation& \bf Description\\
    \hline
    $\mathcal{G}$ & Underlying graph \\
    $\mathcal{V}$ & Set of nodes in $\mathcal{G}$ \\
    $\mathcal{E}$ & Set of edges in $\mathcal{G}$ \\
    $\mathcal{U}$ & Set of users \\
    $\mathcal{I}$ & Set of items \\
    $R$ & user--item rating matrix (interactions)\\
    $\mathcal{M}$ & Set of multimodal information (i.e., textual and visual information) \\
    $L$ & Graph Laplacian of $\mathcal{G}$ \\
    $U$ & Set of eigenvectors of $L$ \\
    $\Lambda$ & Set of eigenvalues of $L$ \\
    $\lambda_{\min}$ & Minimum eigenvalue of $\Lambda$ \\
    $\lambda_{\max}$ & Maximum eigenvalue of $\Lambda$ \\
    $\text{x}$ & Graph signal \\
    $D$ & Diagonal degree matrix \\
    $P$ & Item--item similarity matrix \\
    \bottomrule
\end{tabular}
\label{tab:notations}
\end{table}

Table \ref{tab:notations} summarizes the notation that is used in this paper. This notation will be formally defined in the following sections when we introduce our methodology and technical details.

\section{Related Work}
\label{section 2}
In this section, we review broad research lines related to our study, including 1) GF-based recommendation methods and 2) multimodal recommendation methods.

\textbf{GF-based recommendation.}
In the context of GF, GCN \cite{DBLP:conf/iclr/KipfW17} can be viewed as a parameterized convolutional filter for graphs. A notable GCN-based recommendation method is NGCF \cite{DBLP:conf/sigir/Wang0WFC19}, which was proposed to learn suitable low-pass filters (LPFs) while capturing high-order collaborative signals present in user--item interactions. LightGCN \cite{DBLP:conf/sigir/0001DWLZ020} further demonstrated strong performance by simplifying NGCF, removing both linear transformations and non-linear activations from its GCN layers. However, they are all training-based models that rely on iterative gradient-based optimization over multiple layers, which leads to non-trivial training time and substantial GPU memory consumption, potentially limiting their scalability and ease of deployment in large-scale recommender systems as well as their practicality in latency-sensitive or frequently updated recommendation scenarios.

By bridging the gap between LightGCN and GF approaches, GF-CF \cite{DBLP:conf/cikm/ShenWZSZLL21} was introduced, offering satisfactory recommendation accuracy with minimal computational costs due to its training-free design and a closed-form solution for the infinite-dimensional LightGCN. As a follow-up study, PGSP \cite{DBLP:conf/www/LiuLGL0SG23} employed a mixed-frequency filter that integrates a linear LPF with an ideal LPF. Nevertheless, both GF-CF and PGSP essentially rely on combinations of linear filters and ideal LPFs in the spectral domain, which restricts the flexibility of their frequency responses. In particular, due to the fact that their filters are restricted to either linear or ideal low-pass forms, once a cutoff frequency is chosen, all spectral components above it are uniformly attenuated, leaving limited flexibility to treat different frequency ranges in a more fine-grained manner. In addition, since the ideal LPF component is defined in terms of the eigenvalues and eigenvectors of the graph operator, its construction relies on (exact or approximate) spectral decomposition, which can become a computational bottleneck when the item--item similarity graph is large.

Turbo-CF \cite{DBLP:conf/sigir/ParkSS24} takes a further step toward efficiency by designing polynomial LPFs that retain low-frequency signals without explicitly using an ideal LPF that requires costly matrix decompositions. While this matrix-decomposition-free design substantially reduces runtime, the polynomial filters in Turbo-CF are applied to graph operators whose eigenvalues are not guaranteed to lie within a bounded interval (e.g., $[0,1]$). This mismatch can break the standard assumptions of polynomial graph filtering, making the resulting frequency response harder to analyze and potentially leading to undesirable amplification or attenuation of certain spectral components. In contrast, our proposed \textsf{MM-GF} follows the training-free GF paradigm while explicitly rescaling the graph spectrum so that the eigenvalues fall into a bounded range suitable for polynomial graph filtering. This enables a theoretically grounded design of the filter, preserves desirable low-frequency information, and yields a stable frequency response for multimodal recommendation.

\textbf{Multimodal recommendation.}
Studies on MRSs have been actively conducted to boost the performance of recommender systems by leveraging multimodal features of items (i.e., textual and visual features). Early work such as VBPR \cite{DBLP:conf/aaai/HeM16} extended the Bayesian personalized ranking (BPR) loss \cite{DBLP:journals/corr/abs-1205-2618} by incorporating visual features into matrix factorization, which effectively alleviates item cold-start issues but still models only low-order user–item interactions within a single visual modality. With the rise of graph-based recommendation, GCN-based multimodal models have been proposed to capture high-order connectivity on user–item graphs. For example, GRCN \cite{DBLP:conf/mm/WeiWN0C20} refined the user–item bipartite graph by pruning false-positive edges based on multimodal content, thereby improving robustness to noisy implicit feedback at the cost of an additional graph-refining module and iterative training. LATTICE \cite{DBLP:conf/mm/Zhang00WWW21} explicitly constructed modality-aware item–item graphs and applied graph convolution to capture high-order semantic item relations, which yields strong accuracy but incurs substantial memory and time overhead because the latent item–item graphs scale quadratically with the number of items.

More recent studies introduced self-supervised and structure-regularized graph learning for multimodal recommendation. BM3 \cite{DBLP:conf/www/ZhouZLZMWYJ23} bootstrapped latent representations from user--item interactions and multimodal features using a contrastive self-supervised objective, avoiding negative sampling and auxiliary graphs while still requiring the optimization of a deep GCN backbone and multiple multimodal losses. FREEDOM \cite{DBLP:conf/mm/ZhouS23} simplified LATTICE by freezing the latent item--item graph and incorporating a degree-sensitive edge-pruning strategy on the user--item graph, which improves both recommendation accuracy and memory efficiency on large graphs; however, it remains a fully trainable GNN-based architecture with non-trivial hyperparameter sensitivity (e.g., pruning ratio, loss weights). In MGCN \cite{DBLP:conf/mm/Yu0LB23}, noisy modality features were explicitly purified and enriched in separate graph views before being fused by a behavior-aware module, improving robustness to noisy content at the cost of additional denoising networks and fusion components. Beyond standard bipartite graphs, LGMRec \cite{DBLP:conf/aaai/GuoL0WSR24} constructed a modality-aware hypergraph to jointly model local and global user interests, while PGL \cite{DBLP:conf/aaai/Yu0LB25} extracted principal subgraphs from user--item interactions to preserve individual (high-frequency) structural information. These models enhance expressiveness by introducing hypergraph modules and principal subgraph learning, but their complex architectures and training objectives further increase computational and implementation overhead.

In summary, existing multimodal recommenders can be grouped into 1) shallow CF models with side multimodal features, and 2) training-based graph models that learn high-capacity GNNs, auxiliary graphs, or self-supervised objectives. While the latter achieve strong accuracy, they typically require expensive end-to-end training, careful negative sampling or contrastive loss design, and additional graph-learning modules. In contrast, our proposed \textsf{MM-GF} takes a complementary, training-free approach: it models multimodal recommendation as spectral polynomial graph filtering, using closed-form filters defined on preconstructed user--item and item--item graphs. By avoiding parameterized GNN layers and negative-sampling-based optimization, \textsf{MM-GF} maintains the ability to exploit high-order collaborative and multimodal signals, while enjoying lower computational complexity and a simpler, model-agnostic deployment pipeline.

\section{Preliminaries}
\label{section 3}

We provide some preliminaries such as the GF mechanism and the problem definition.

\subsection{Notion of GF}
We provide fundamental principles of GF (or equivalently, graph signal processing) \cite{DBLP:journals/pieee/OrtegaFKMV18, DBLP:conf/cikm/ShenWZSZLL21, DBLP:conf/sigir/ParkSS24}. First, we consider an undirected graph $\mathcal {G=(V,E)}$, which is represented by an adjacency matrix $A$ that indicates the presence or absence of edges between nodes. The Laplacian matrix $L$ of $\mathcal G$ is defined as $L=D-A$, where $D$ is the degree matrix of $A$ \cite{DBLP:journals/siamdm/GroneM94}. Meanwhile, a graph signal on $\mathcal G$ is expressed as $\mathbf{x} \in \mathbb R^{|\mathcal V|}$, where each $x_i$ indicates the signal strength at the corresponding node $i$ in ${\bf x}$. The smoothness of a graph signal $\mathbf{x}$ can be mathematically measured by
\begin{equation}
    S(\mathbf{x}) = \sum_{i,j}A_{ij}(x_i-x_j)^2 = \mathbf{x}^T L\mathbf{x},
\end{equation}
where a smaller $\frac{S(\mathbf{x})}{\|\mathbf{x}\|_2}$ indicates a smoother $\mathbf{x}$.

The graph Fourier transform (GFT) converts a graph signal into the frequency domain using the eigenvectors of the graph Laplacian $L=U\Lambda U^T$, where $U\in\mathbb R^{|V| \times |V|}$ is the matrix whose columns correspond to a set of eigenvectors of $L$ and $\Lambda$ is a diagonal matrix containing the set of eigenvalues of $L$. Thus, the graph signal $\mathbf{x}$ can be transformed into $\hat {\mathbf{x}} = U^T \mathbf{x}$, which utilizes the spectral characteristics of the underlying graph to examine the latent structure of the graph signal. The GFT is used to perform \textit{graph convolution}, with the aid of {\it graph filters}, which is mathematically defined as follows.

\textbf{Definition 1} (Graph filter). The graph filter $H(L)$ is defined as
\begin{equation}
H(L) = U \text{diag}(h(\lambda_1), \dots, h(\lambda_{|\mathcal{V}|})) U^T,
\end{equation}
where $h()$ is the frequency response function that maps eigenvalues $\{\lambda_1,\dots,\lambda_{|\mathcal{V}|}\}$ of $L$ to $\{h(\lambda_1),\dots,h(\lambda_{|\mathcal{V}|})\}$.

\textbf{Definition 2} (Graph convolution). The graph convolution of a signal $\mathbf{x}$ and a graph filter $H(L)$ is defined as
\begin{equation}
    H(L) \mathbf{x} = U \text{diag}(h(\lambda_1), \dots, h(\lambda_{|\mathcal{V}|})) U^T\mathbf{x}.
\end{equation}

\subsection{Problem Definition} We formally present the problem of top-$K$ multimodal recommendations. First, let $\mathcal U $ and $\mathcal I$ denote the set of users and the set of items, respectively. A user--item rating matrix is denoted as $R \in \mathbb{R}^{|\mathcal U | \times |\mathcal I |}$, and $\mathcal M$ is the set of multiple modalities. In this paper, we use the textual and visual features denoted as $\mathcal M = \{\text{txt} \, , \text{img}\}$. We denote the feature matrix of modality $m \in \mathcal{M}$ as $X^m \in \mathbb R ^{|\mathcal I | \times d_m}$ for the dimensionality $d_m$. The objective of the multimodal recommendation task is to recommend top-$K$ items to each user using multimodal features of items as well as user--item interactions representing the ratings.

\section{Methodology}
\label{section 4}

In this section, we elaborate on the proposed \textsf{MM-GF} method and its computational complexity as well as our research motivation.

\subsection{Motivation and Challenges}
\label{4.1.Motivation}
Conventional training-free GF-based recommendation methods \cite{DBLP:conf/cikm/ShenWZSZLL21,DBLP:conf/www/LiuLGL0SG23} start by constructing a graph structure, where nodes represent items and edges correspond to item--item similarities. The graph construction process is formulated as follows:
\begin{equation}
\label{conventional_graph}
\begin{aligned}
    \tilde{P} = \tilde{R}^T\tilde{R} ; \tilde{R} = D^{-1/2}_rRD^{-1/2}_c,
\end{aligned}
\end{equation}
where $\tilde{R}$ is the normalized rating matrix; $D_r=\text{diag}(R\mathbf{1})$ and $D_c = \text{diag} (\mathbf{1}^TR)$ are the diagonal degree matrices of users and items, respectively, for the all-ones vector ${\mathbf 1}$; and $\tilde{P}$ is the adjacency matrix of the item--item similarity graph.

Recently, \cite{DBLP:conf/sigir/ParkSS24, DBLP:conf/www/ParkYS25, DBLP:conf/www/KimCPS25} utilized asymmetric normalization on $R$ to regularize the popularity of users or items before calculating $\tilde P$, which is formulated as follows:
\begin{equation}
\label{asymmetric_normalization_on_R}
\begin{aligned}
    \tilde{P}=\tilde{R}^T\tilde{R} ; \tilde R = D_r^{-\alpha}RD_c^{\alpha-1},
\end{aligned}
\end{equation}
where $\alpha \in [0,1]$ is the hyperparameter to control the normalization among users and items. Intuitively, as $\alpha$ decreases, the influence of popular items on $\tilde R$ becomes weak. Additionally, to address over-smoothing or under-smoothing problems, an adjustment hyperparameter was adopted for the graph $\bar P$ as:
\begin{equation}
\label{Adjustment_for_P}
\begin{aligned}
    \bar{P}=\tilde P^{\circ s},
\end{aligned}
\end{equation}
where $s$ is an adjustment parameter tunable on the validation set.

\begin{figure}[t]
    \centering
    \begin{subfigure}{0.48\textwidth}
        \centering
        \includegraphics[width=\linewidth]{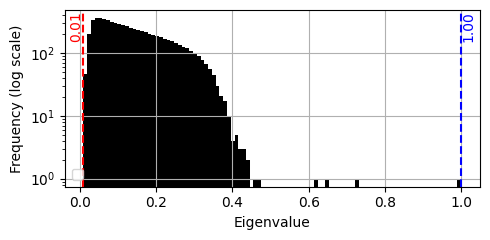}
        \caption{$\alpha=0.5, s=1$}
        \label{fig:histogram of P(GF-CF)}
    \end{subfigure}
    \hfill
    \begin{subfigure}{0.48\textwidth}
        \centering
        \includegraphics[width=\linewidth]{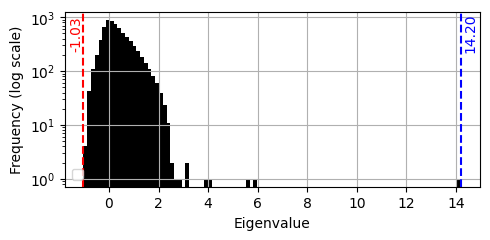}
        \caption{$\alpha=0.7, s=0.6$}
        \label{fig:histogram of P(Turbo-CF)}
    \end{subfigure}
    \begin{subfigure}{0.48\textwidth}
        \centering
        \includegraphics[width=\linewidth]{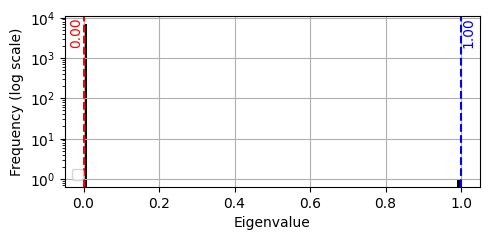}
        \caption{$\tilde P_\text{txt}$ for $\alpha=0.5, s=1$}
        \label{fig:histogram of P_text(GF-CF)}
    \end{subfigure}
    \hfill
    \begin{subfigure}{0.48\textwidth}
        \centering
        \includegraphics[width=\linewidth]{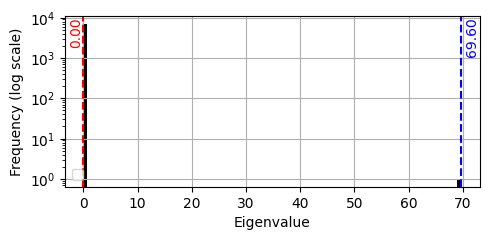}
        \caption{$\bar P_\text{txt}$ for $\alpha=0.7, s=0.6$}
        \label{fig:histogram of P_text(Turbo-CF)}
    \end{subfigure}
    \begin{subfigure}{0.48\textwidth}
        \centering
        \includegraphics[width=\linewidth]{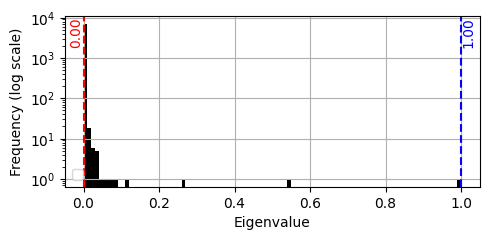}
        \caption{$\tilde P_\text{img}$ for $\alpha=0.5, s=1$}
        \label{fig:histogram of P_img(GF-CF)}
    \end{subfigure}
    \hfill
    \begin{subfigure}{0.48\textwidth}
        \centering
        \includegraphics[width=\linewidth]{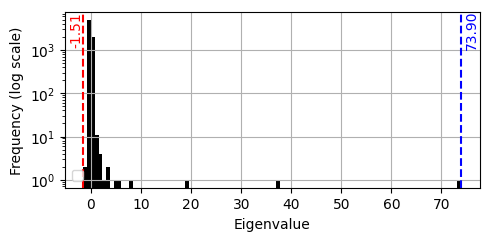}
        \caption{$\bar P_\text{img}$ for $\alpha=0.7, s=0.6$}
        \label{fig:histogram of P_img(Turbo-CF)}
    \end{subfigure}
    \caption{Histogram of eigenvalues for item--item similarity graphs constructed from user--item interactions and multimodal information on the Baby dataset. The left panels show the case of $\alpha=0.5$ and $ s=1$, corresponding to symmetric normalization, while the right panels show the case of $\alpha=0.7$ and $s=0.6$, representing asymmetric normalization with an additional adjustment process. Red dashed lines indicate the minimum eigenvalue of each graph, while blue dashed lines indicate the maximum eigenvalue.}
    \label{fig:all eigenvalue histogram}
    \vspace{-4mm}
\end{figure}

However, as critical issues, introducing two hyperparameters (i.e., $\alpha$ and $s$) may alter the spectral characteristics of the underlying graph. Specifically, the work in \cite{DBLP:conf/cikm/ShenWZSZLL21} proved that all the eigenvalues of $\tilde P = \tilde R^T \tilde R$ where $\tilde R = D_r^{-0.5}R D_c^{-0.5}$ are in $[0,1]$. Nonetheless, as long as parameters $\alpha$ and $s$ are concerned, one cannot always assure that $\bar P$ in Eqs. \eqref{asymmetric_normalization_on_R} and \eqref{Adjustment_for_P} satisfies the aforementioned condition in Section \ref{section1.1} (i.e., the unit interval of eigenvalues) when $\alpha \neq 0.5$ and/or $ s\neq 1$. In Figure \ref{fig:all eigenvalue histogram}, we present two cases: Case 1) $\alpha=0.5$ and $s=1$, corresponding to symmetric normalization, and Case 2) $\alpha=0.7$ and $s=0.6$, corresponding to asymmetric normalization with an additional adjustment process. The observations from this figure pose practical challenges below.

\begin{itemize}
    \item \textbf{User--item interactions.} In Figures \ref{fig:histogram of P(GF-CF)} and \ref{fig:histogram of P(Turbo-CF)}, each histogram illustrates the distribution of eigenvalues of two distinct item--item similarity graphs constructed by Case 1 and 2, respectively, for user--item interaction data on the Baby dataset. We observe from Figure \ref{fig:histogram of P(GF-CF)} that all eigenvalues lie within the interval $[0,1]$ for $\alpha=0.5$ and $s=1$ whereas, for another setting (i.e., $\alpha = 0.7$ and $s=0.6$) in Figure \ref{fig:histogram of P(Turbo-CF)}, they range approximately from -1 to 14, which violates the unit-interval condition of the eigenvalues.

    \item \textbf{Multimodal information.} Figures \ref{fig:histogram of P_text(GF-CF)}--\ref{fig:histogram of P_img(Turbo-CF)} present histograms of eigenvalue distributions for the Baby dataset. Figures \ref{fig:histogram of P_text(GF-CF)} and \ref{fig:histogram of P_img(GF-CF)} illustrate Case 1, and Figures \ref{fig:histogram of P_text(Turbo-CF)} and \ref{fig:histogram of P_img(Turbo-CF)} illustrate Case 2, each representing the eigenvalue distribution of the corresponding item–item similarity graph in the multimodal graph structure. The maximum eigenvalues of $\bar P_{\text{txt}}$ and $\bar P_{\text{img}}$ are approximately 70, as illustrated in Figures \ref{fig:histogram of P_text(Turbo-CF)} and \ref{fig:histogram of P_img(Turbo-CF)}, which is considerably larger than the case of user--item interactions. The use of such unbounded frequency components not only hinders the effectiveness of polynomial graph filters but also introduces a non-ideal outcome when integrating multiple graph filters.
\end{itemize}

As observed in Figure \ref{fig:all eigenvalue histogram}, when we construct the graph operators exactly as in previous GF-based methods \cite{DBLP:conf/sigir/ParkSS24, DBLP:conf/www/ParkYS25, DBLP:conf/www/KimCPS25}, the resulting eigenvalue spectrum deviates substantially from the canonical interval $[0, 1]$, with plenty of eigenvalues lying far below 0 or far above 1; this mismatch becomes problematic precisely when the polynomial LPFs from prior studies are na\"ively applied to such spectra, since they were originally derived under the assumption of eigenvalues confined to $[0, 1]$. Symmetric normalization in \cite{DBLP:conf/cikm/ShenWZSZLL21, DBLP:conf/www/LiuLGL0SG23, DBLP:conf/www/XiaLGLLG22} typically follow that the spectrum lies within $[0, 1]$, which was proved by \cite{DBLP:conf/cikm/ShenWZSZLL21}. When eigenvalues are mapped out of this range, the polynomial filter may yield inaccurate frequency responses, deviating from its designed frequency characteristics. Moreover, as the polynomial order increases, the filter response for large-magnitude eigenvalues can grow dramatically; this results in a disproportionate emphasis on certain frequency components, thereby degrading the model performance. For example, consider a polynomial LPF $h(\lambda) = 1 - \lambda^2$. In a bounded frequency range of $[0,1]$, filtered values close to 0 correspond to high-frequency components, whereas those close to 1 correspond to low-frequency components. However, if the spectrum contains eigenvalues such as $\lambda_1 \simeq -1$ and $\lambda_2 \simeq 14$, as shown in Figure \ref{fig:histogram of P(Turbo-CF)}, then their corresponding filter responses become $h(\lambda_1) \simeq 0$ and $h(\lambda_2) \simeq -195$, respectively. This phenomenon results in signals being assigned to anomalous frequencies instead of the targeted ones. Furthermore, this problem becomes particularly critical for multimodal information, where signals are more susceptible to distortion compared to the case of user--item interactions. For example, if the spectrum contains an eigenvalue $\lambda_3 \simeq 70$, as illustrated in Figures \ref{fig:histogram of P_text(Turbo-CF)} and \ref{fig:histogram of P_img(Turbo-CF)}, then its corresponding filter response becomes $h(\lambda_3) \simeq -4899$, leading to far more severe distortion than the case of user--item interactions. Ultimately, since MRSs require integrating signals across multiple modalities, aggregating such severely distorted signals can undermine the overall reliability of the recommendation process.

Additionally, despite the extensive literature on CF, there currently exists no training-free popularity-balancing formulation that preserves the spectral properties required for GF. Prior approaches addressing popularity bias (e.g., \cite{DBLP:conf/sigir/0002YKL25, DBLP:journals/tors/PengSLM25}) rely on learning-based adjustments and therefore cannot be applied when an analytically grounded, training-free mechanism is desired. This limitation further motivates the need for a principled normalization strategy, which our framework is designed to provide.

Addressing a different issue, while recent GF models \cite{DBLP:conf/sigir/ParkSS24,DBLP:conf/www/ParkYS25, DBLP:conf/www/KimCPS25} demonstrated that a polynomial graph filter can be implemented as the form of $\sum_{k=1}^K a_k \bar{P}^k$, their empirical application was confined only to specific filters. This limited scope highlights a gap in the broader applicability of the GF methodology. Moreover, it is of paramount importance to effectively aggregate information across multiple modalities for accurate multimodal recommendations. In light of these challenges, a key question arises: "How can we design an efficient and effective GF method for multimodal recommendations by not only addressing the issue of filter design but also maximally exploiting heterogeneous characteristics across modalities?" To answer this question, we will outline the proposed \textsf{MM-GF} method tailored for multimodal recommendations in the following subsection.

\begin{figure}[t]
    \centering
    \includegraphics[width=\columnwidth]{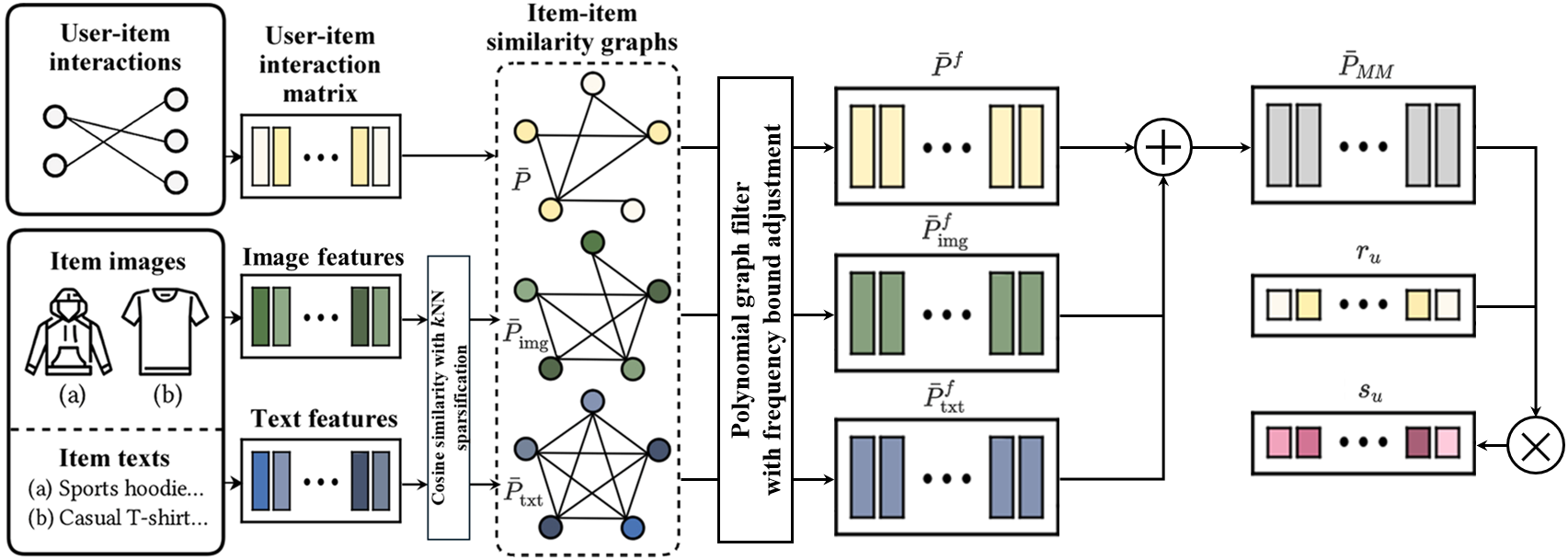}
    \vspace{-1.5mm}
    \caption{The schematic overview of \textsf{MM-GF}.}
    \vspace{-3mm}
    \label{overview}
\end{figure}

\subsection{Proposed Method: \textsf{MM-GF}}

In this subsection, we describe the graph construction process and the filter design process in \textsf{MM-GF}. The schematic overview of \textsf{MM-GF} is illustrated in Figure \ref{overview}. We provide the pseudocode of \textsf{MM-GF} in Algorithm 1, which summarizes the whole process.

\subsubsection{Graph Construction for User--Item Interactions}
Similarly as in existing GF-based recommendation methods \cite{DBLP:conf/sigir/ParkSS24, DBLP:conf/www/ParkYS25, DBLP:conf/www/KimCPS25}, the item--item similarity graph for user--item interactions is constructed as
\begin{equation}
\label{degree_normalization}
\tilde{P} = \tilde{R}^T \tilde{R}; \quad \tilde{R} = D_r^{-\alpha} R D_c^{\alpha-1},
\end{equation}
where $\alpha$ is a hyperparameter controlling the asymmetric normalization along users and items \cite{DBLP:conf/sigir/ParkSS24, DBLP:conf/www/ParkYS25, DBLP:conf/www/KimCPS25}. Additionally, we adjust $\tilde{P}$ using the Hadamard power $\bar{P} = \tilde{P}^{\circ s}$ as properly adjusting the item–item similarity graph using the Hadamard power was shown to produce more accurate recommendations \cite{DBLP:conf/sigir/ParkSS24, DBLP:conf/www/ParkYS25, DBLP:conf/www/KimCPS25}, where $\circ$ denotes the Hadamard (element-wise) power and $s$ is a filter adjustment hyperparameter.

\subsubsection{Graph Construction for Textual and Visual Modalities}
\label{graph construction}

\begin{algorithm}[!t]
\caption{\textsf{MM-GF}}
\label{alg: MM-GF}
\begin{algorithmic}[1]
\REQUIRE User--item rating matrix $R$, textual feature matrix $X^{\text{txt}}$, visual feature matrix $X^{\text{img}}$
\ENSURE Asymmetric normalization along users/items: $\alpha$, filter adjustment: $s$, balancing factors among the three item--item similarity graphs: $\beta, \gamma$
\STATE $\tilde R \gets D_r^{-\alpha}RD_c^{\alpha -1}$ where $D_r=\text{diag}(R\mathbf{1})$ and $D_c = \text{diag}(\mathbf{1}^TR)$
\STATE $\tilde P \gets \tilde R ^T \tilde R$
\STATE $\bar P \gets \tilde P^{\circ s}$
\FOR {$m\in\{\text{txt}, \text{img}\}$}
    \STATE $S^m\gets CS(X^m)$ where $CS$ is cosine similarity
    \STATE $\hat S^m \gets \text{top-}k(S^m)$ where $\text{top-}k$ is top-$k$ preservation
    \STATE $\tilde S_m \gets D_{m,r}^{-\alpha}\hat{S}^mD_{m,c}^{\alpha -1}$ where $D_{m,r}=\text{diag}(\hat S^m\mathbf{1})$ and $D_{m,c} = \text{diag}(\mathbf{1}^T \hat S^m)$
    \STATE $\tilde P_m \gets \tilde S_m \tilde S_m^T$
    \STATE $\bar P_m \gets \tilde P_m^{\circ s}$
\ENDFOR
\FOR {$\bar P_\star \in \{\bar P,\bar P_\text{txt},\bar P_\text{img}\}$}
    \STATE $\bar P_\star^f\gets \sum_{k=1}^K\frac{a_k}{(\lambda^\star)^{k-1}}(\bar P_\star - \lambda_{min}^\star  I)^k$ where $\lambda_{min}^\star$ is minimum eigenvalue of $\bar P_\star$
\ENDFOR
\STATE $\bar P_{\text{MM}} \gets \bar P^f+ \beta \bar P^f_{\text{txt}} + \gamma \bar P^f_{\text{img}}$
\STATE $s_u \gets r_u \bar P_{\text{MM}}$
\RETURN $s_u$

\end{algorithmic}
\end{algorithm}

\begin{figure}[t]
    \centering
    \begin{subfigure}{0.47\textwidth}
        \centering
        \includegraphics[width=\linewidth]{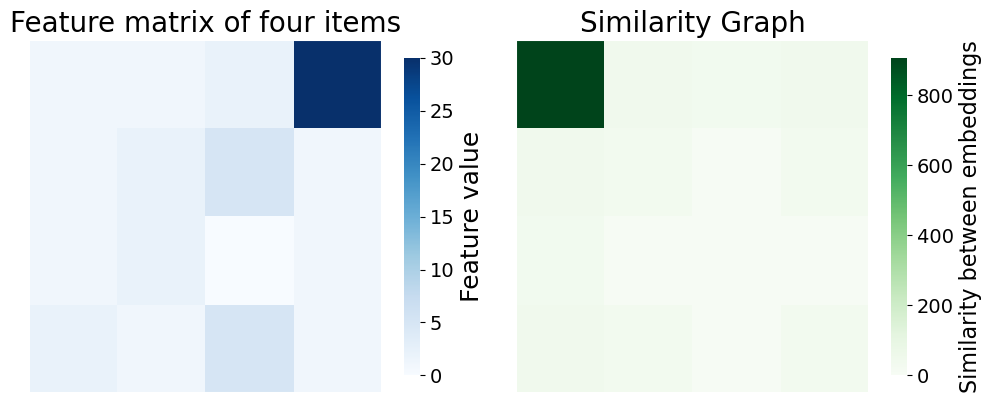}
        \caption{Outliers}
        \label{fig:outlier}
    \end{subfigure}
    \hfill
    \begin{subfigure}{0.475\textwidth}
        \centering
        \includegraphics[width=\linewidth]{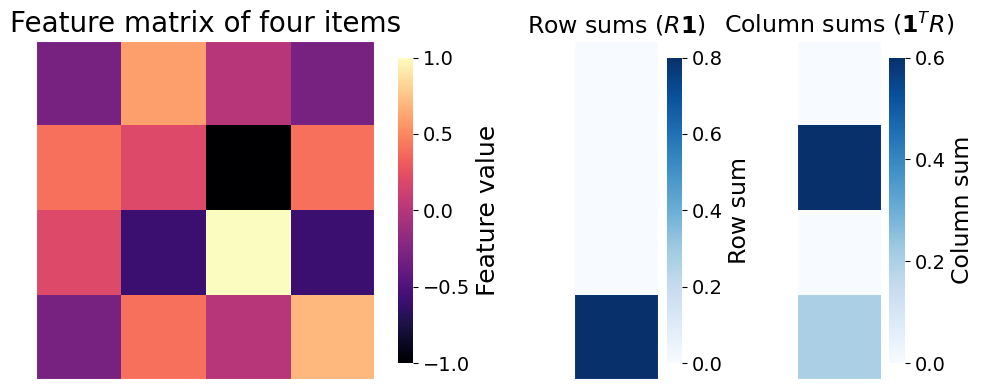}
        \caption{Singularities}
        \label{fig:singularity}
    \end{subfigure}
    \caption{The impact of outliers and negative values (leading to singularities).}
    \label{fig:both}
    \vspace{-4mm}
\end{figure}

If we na\"ively apply the same graph construction strategy used for user–item interactions to the textual and visual modalities, then we may fail due to the existence of {\it outliers} or {\it negative values} in feature vectors generated through sentence-transformers \cite{DBLP:conf/emnlp/ReimersG19} and pre-trained convolutional neural networks (CNNs) \cite{DBLP:conf/www/HeM16}. As illustrated in Figure \ref{fig:both}, constructing item--item similarity graphs for multimodal information using this strategy can lead to two critical issues. First, outliers in the feature vectors may cause certain elements of the graph to take excessively large values, thereby dominating the similarity graph and suppressing the contribution of other similarities, which are then forced to remain close to zero (see Figure \ref{fig:outlier}). This distortion makes it difficult to accurately capture the true relationships between items. Second, when feature vectors include both positive and negative values, the row and column sums in the normalization process can approach zero (see Figure \ref{fig:singularity}). Dividing by such near-zero values results in singularities in the normalized graph, ultimately leading to instability and unreliable graph structures. Therefore, it is necessary to construct a graph structure in the form of similarity graphs tailored for multimodal information. While there exist a variety of similarity calculation strategies such as the Pearson correlation coefficient \cite{cohen2009pearson} and the Gaussian kernel \cite{DBLP:journals/spm/ShumanNFOV13}, we empirically found that cosine similarity-based calculation guarantees superior performance on benchmark datasets in terms of recommendation accuracy consistently. In this context, we adopt cosine similarity to construct item--item similarity graphs unless otherwise stated.

To be specific, following \cite{DBLP:conf/mm/Zhang00WWW21, DBLP:conf/mm/ZhouS23, DBLP:conf/mm/Yu0LB23}, \textsf{MM-GF} constructs an item--item similarity graph by applying $k$NN sparsification on raw features corresponding to each modality $m$. Given $N$ items, the similarity score $S_{i,j}^m$ between item $i$ and $j$ is computed using cosine similarity based on their raw modality-$m$ feature vectors $(x_i^m, x_j^m)$, as follows:

\begin{equation}
\label{cosine_similarity}
S_{i,j}^m=\frac{x_i^m  (x_j^m)^T}{\Vert x_i^m\Vert\, \Vert x_j^m\Vert}.
\end{equation}

Here, $S_{i,j}^m$ represents the element located at the $i$-th row and $j$-th column of the similarity matrix $S^m \in \mathbb R^{N\times N}$.
To sparsify the graph, we apply $k$-nearest neighbor preservation in \cite{DBLP:conf/mm/ZhouS23} to convert the weighted matrix $S^m$ into an unweighted version. For each item $i$, only its top-$k$ most similar neighbors are kept, resulting in the binary matrix $\hat S^m$ defined by:

\begin{equation}
\label{binary_similarity}
\hat S_{i,j}^m =
    \begin{cases}
        1, & \text{if } S^m_{i,j} \in \text{top-}k(S^m_i), \\
0, & \text{otherwise},
    \end{cases}
\end{equation}
where $\hat S ^m$ is the binarized matrix. Finally, the resulting feature matrix $\hat S^m$ is then normalized to construct the item--item similarity graph $\tilde{P}_m$ for modality $m$:
\begin{equation}
\tilde{P}_m = \tilde{S}_m \tilde{S}_m^T; \quad \tilde{S}_m = D_{m,\text{r}}^{-\alpha} \hat{S}^m D_{m,\text{c}}^{\alpha-1},
\end{equation}
where $D_{m,\text{r}} = \operatorname{diag}(\hat S^m \mathbf{1})$ and $D_{m,\text{c}} = \operatorname{diag}(\mathbf{1}^T \hat S^m)$ are the diagonal degree matrices for modality $m$. Likewise, we adjust $\tilde{P}_m$ as $\bar{P}_m = \tilde{P}_m^{\circ s}$ in the graph construction process.\footnote{While using different $\alpha$'s and $s$'s for each modality certainly increases recommendation accuracy, we use the same values of $\alpha$ and $s$ as those for user--item interactions across multiple modalities for simplicity.} Consequently, we construct item–-item similarity graphs for the textual and visual modalities, denoted by $\bar{P}_{\text{txt}}$ and $\bar{P}_{\text{img}}$, respectively.

\subsubsection{Filter Design with Multiple Modalities}
Next, we are interested in how to precisely adjust the frequency bound of item--item similarity graphs. Specifically, when we adopt polynomial graph filters in \cite{DBLP:conf/sigir/ParkSS24, DBLP:conf/www/ParkYS25, DBLP:conf/www/KimCPS25}, it is necessary that all eigenvalues of the item--item similarity graph lie within $[0, 1]$. This approach defines a polynomial graph filter as $\sum_{k=1}^{K} a_k \bar P_\star^k$, which corresponds to the frequency response function $h(\lambda) = \sum_{k=1}^K a_k (1 - \lambda)^k$. However, as we observed in Figure ~\ref{fig:all eigenvalue histogram} of Section \ref{4.1.Motivation}, certain eigenvalues exceed the specified bounds.

To address the issue, we propose a more appropriate and technically valid polynomial graph filter by replacing $\bar P_\star$ with $\bar P_\star - \lambda_{min}^\star I$ and further normalizing each term by $(\lambda^\star)^{k-1}$ to control the spectral scaling, where $\bar P_{\star} \in \{\bar P,\bar P_{\text{txt}},\bar P_{\text{img}}\}$ and $\lambda^{\star}_{min}$ is the minimum eigenvalue of $\bar P_{\star}$. This ensures that all eigenvalues of $\bar P_\star - \lambda_{min}^\star I$ lie within the range $[0, \lambda^\star]$ where $\lambda^{\star}=\lambda_{max}^{\star}-\lambda_{min}^{\star}$ and $\lambda_{min}^{\star}$ is the minimum eigenvalue of $\bar P_{\star}$.\footnote{$\lambda_{max}^\star$ and $\lambda_{min}^\star$ are obtained by solving the characteristic equation $p(\lambda) = \text{det}(\bar P_\star -\lambda  I)$.} To this end, we would like to establish the following theorem, which characterizes the frequency response function of a polynomial graph filter that accommodates the bounded range of eigenvalues.

\begin{theorem}
    \label{thm_poly}
     For any admissible range of eigenvalues, the matrix polynomial can be expressed as:

     \begin{equation}
     \label{poly_filter}
         \bar P ^f_\star = \sum_{k=1}^{K} \frac{a_k}{(\lambda^\star)^{k-1}} (\bar{P}_\star-\lambda_{min}^\star  I)^k,
     \end{equation}
     where $\bar P^f_\star$ is the graph filter associated with the similarity graph $\bar{P}_\star$, having the frequency response function of $h(\lambda)=\sum_{k=1}^K\frac{a_k}{(\lambda^\star)^{k-1}} (\lambda^\star-\lambda)^k$, and $\lambda^\star = \lambda^\star_{max} - \lambda^\star_{min}$. Here, $\lambda^\star_{min}$ and $\lambda^\star_{min}$ are the minimum and maximum eigenvalues of $\bar P_\star$, respectively.
\end{theorem}

\begin{pf}
    First, we define the graph Laplacian $L_\star$ for a specific modality $\star$ as follows:

    \begin{equation}
        \bar L_\star = \lambda_{max}^\star I - \bar P_\star = U\Lambda^\star U^T,
    \end{equation}
    where $\Lambda^\star$ is a set of eigenvalues of $L_\star$. Since the eigenvalues of $\bar P_\star$ fall within $[\lambda_{min}^\star, \lambda_{max}^\star]$, those of $L_\star$ lie within $[0, \lambda^\star]$. Based on the graph Laplacian $L_\star$, we have

    \begin{equation}
        \begin{aligned}
            \bar P_\star - \lambda^\star_{min}I &= \lambda^\star_{max}I-L_\star - \lambda^\star_{min}I \\ &= U(\lambda^\star_{max}I - \Lambda^\star - \lambda^\star_{min}I)U^T \\ &= U(\lambda^\star I -\Lambda^\star)U^T,
        \end{aligned}
    \end{equation}
    which implies that $\bar P_\star - \lambda^\star_{min}I$ is a graph filter having a frequency response function of $h(\lambda)=\lambda^\star-\lambda$ for $[0, \lambda^\star]$. This corresponds to a linear LPF whose frequency response defined over the interval $[0,\lambda^\star]$. Now, we consider the $k$th-order polynomial filter. Then, we have

    \begin{equation}
        \begin{aligned}
            \frac{1}{(\lambda^\star)^{k-1}}(\bar P_\star-\lambda^\star_{min}I)^k &=\frac{1}{(\lambda^\star)^{k-1}} \underbrace {U (\lambda^\star I-\Lambda^\star) U^T \dots U(\lambda^\star I-\Lambda^\star) U^T}_{k\ \text{times}} \\&=U\left( \frac{(\lambda^\star I -\Lambda^\star)^k}{(\lambda^\star)^{k-1}} \right) U^T,
        \end{aligned}
    \end{equation}
    which indicates a graph filter with a frequency response function of $h(\lambda) = \frac{(\lambda^\star - \lambda)^k}{(\lambda^\star)^{k-1}}$ for $\lambda\in[0, \lambda^\star]$. Since $h(0) = \lambda^\star$, $h(\lambda^\star)=0$, and $h(\lambda)$ is a monotonically decreasing function in the interval $[0, \lambda^\star]$, the filtered value is still bounded in $[0,\lambda^\star]$. Therefore, the proposed filter design guarantees spectral boundedness of $\bar{P}_\star^f$ within $[0, \lambda^\star]$, even when eigenvalues of $\bar{P}_\star$ exceed the standard unit interval. This completes the proof of this theorem.
    \hfill $\square$
\end{pf}

Intuitively, Eq. \eqref{poly_filter} shifts and rescales the spectrum of $\bar P_{\star}$ so that all eigenvalues fall into the interval $[0,\lambda^{\star}]$, and then applies a polynomial graph filter on this transformed spectrum, thereby resolving the instability that occurs when conventional polynomial graph filters are directly applied to matrices whose eigenvalues lie outside the assumed bounded range.

While directly specifying the coefficients $a_k$ in Eq. \eqref{poly_filter} provides precise, analytically designed control over the frequency response, it also limits expressiveness because the filter shape is fixed a priori by the designer. In practice, since it is often difficult to know in advance which frequency components should be emphasized or attenuated for a given dataset, a hand-crafted filter may fail to align with the true spectral characteristics of real-world recommendation graphs. To address this, we treat $a_k$ as tunable hyperparameters, allowing data-driven selection via validation rather than enforcing a single fixed analytic form.

Finally, \textsf{MM-GF} optimally fuses multimodal information as a weighted sum of the graph filters across different modalities:
\begin{equation}
\label{beta_gamma}
\bar{P}_{\text{MM}} = \bar{P}^f + \beta \bar{P}_{\text{txt}}^f + \gamma \bar{P}_{\text{img}}^f,
\end{equation}
where $\beta$ and $\gamma$ are hyperparameters balancing among the three similarity graphs. The filtered signal for user $u$ (i.e., the predicted preference scores of user $u$) is finally given by
\begin{equation}
s_u = r_u \bar{P}_{\text{MM}},
\end{equation}
where ${\mathbf{r}}_{u}$ denotes the $u$-th row of $R$ and is utilized as the graph signal for user $u$.

\subsection{Computational Complexity Analysis}

Our proposed method leverages training-free graph filtering with a polynomial filter design and spectral components derived from the item--item graph. Here, we analyze the computational complexity of the proposed \textsf{MM-GF} method from three viewpoints.

\paragraph{Polynomial graph filtering.} The polynomial graph filter of order $K$ is implemented via $K$ matrix-vector multiplications. Given an $N \times N$ item-item graph (where $N$ is the number of items), each multiplication has a complexity of $\mathcal{O}(N^2)$. Thus, the overall filtering step has a time complexity of $\mathcal{O}(KN^2)$, which is efficient for moderate values of $K$.

\paragraph{Spectral component computation.} In our implementation, we use exact eigendecomposition to obtain the graph Laplacian's spectrum, which has a computational complexity of $\mathcal{O}(N^3)$. However, it is important to emphasize that in our experiments—conducted on standard multimodal recommendation benchmarks—the graph sizes are moderate, and the eigendecomposition step does not dominate the overall runtime, accounting for only a small fraction of the total computation. Hence, the runtime is practically acceptable, as also supported by our empirical timing results (see Table \ref{runtime_table}).

\paragraph{Scalability via approximation.} For large-scale applications where full eigendecomposition may become infeasible, our framework remains applicable by relying on approximate methods. Specifically, since the polynomial filter only requires the largest and smallest eigenvalues of the Laplacian, we can employ efficient approximation techniques such as the Lanczos algorithm \cite{cullum2002lanczos} to estimate these extremal eigenvalues with significantly lower complexity than $\mathcal{O}(N^3)$ \cite{arora2005fast}. Thus, our method maintains scalability without compromising its training-free and model-agnostic design.

\section{Experimental Results and Analyses}
\label{section 5}

In this section, we systematically conduct extensive experiments to answer the following key five research questions (RQs).

\begin{itemize}
    \item \textbf{RQ1}: How does \textsf{MM-GF} perform compared with the state-of-the-art multimodal recommendation methods? 
    \item \textbf{RQ2}: How efficient is \textsf{MM-GF} in terms of runtime and scalability for multimodal recommendations? 
    \item \textbf{RQ3}: How does each component of \textsf{MM-GF} affect its recommendation accuracy? 
    \item \textbf{RQ4}: How does \textsf{MM-GF} perform in cold-start settings? 
    \item \textbf{RQ5}: How sensitive is \textsf{MM-GF} under noisy multimodal data settings? 
    \item \textbf{RQ6}: How does the polynomial graph filter of \textsf{MM-GF} operate in the spectral domain?
    \item \textbf{RQ7}: How sensitive is \textsf{MM-GF} to variations in its hyperparameters?
\end{itemize}

\subsection{Experimental Settings}

\noindent\textbf{Datasets.} We use three widely used benchmark datasets in recent MRSs \cite{DBLP:conf/mm/Zhang00WWW21, DBLP:conf/mm/Yu0LB23, DBLP:conf/mm/ZhouS23, zhou2023comprehensive}, which were collected from Amazon \cite{DBLP:conf/sigir/McAuleyTSH15}: (a) Baby, (b) Sports and Outdoors, and (c) Clothing, Shoes, and Jewelry, which we refer to as Baby, Sports, and Clothing, respectively, in brief.\footnote{Datasets are officially available at https://jmcauley.ucsd.edu/data/amazon/links.html.} The above three datasets contain textual and visual features as well as user--item interactions. For the textual modality, we use 384-dimensional textual embeddings by combining the title, descriptions, categories, and brand of each item and adopt sentence-transformers \cite{DBLP:conf/emnlp/ReimersG19}. For the visual modality, we use 4,096-dimensional visual embeddings obtained by applying pre-trained CNNs \cite{DBLP:conf/www/HeM16}. Table \ref{table:datasets} provides a summary of the statistics for each dataset.

\begin{table}[!t]
\footnotesize
  \captionsetup{skip=2pt}
  \caption{The statistics of three benchmark datasets.}
  \begin{tabular}{ccccccl}
    \toprule
    Dataset & \# of users & \# of items & \# of interactions & Sparsity \\
    \midrule
    Baby & 19,445 & 7,050 & 160,792 & 99.88\%\\
    Sports & 35,598 & 18,357 & 296,337 & 99.95\%\\
    Clothing & 39,387 & 23,033 & 278,677 & 99.97\% \\
  \bottomrule
\end{tabular}
\vspace{-2mm}
\label{table:datasets}
\end{table}

\noindent\textbf{Competitors.} To validate the effectiveness of \textsf{MM-GF}, we conduct a comparative analysis with seven state-of-the-art recommendation methods, especially those built upon neural network models including GCNs. The benchmark methods include not only a single-modal recommendation method (LightGCN \cite{DBLP:conf/sigir/0001DWLZ020}) but also multimodal recommendation methods (VBPR \cite{DBLP:conf/aaai/HeM16}, GRCN \cite{DBLP:conf/mm/WeiWN0C20}, LATTICE \cite{DBLP:conf/mm/Zhang00WWW21}, BM3 \cite{DBLP:conf/www/ZhouZLZMWYJ23}, FREEDOM \cite{DBLP:conf/mm/ZhouS23}, and MGCN \cite{DBLP:conf/mm/Yu0LB23}).

\noindent\textbf{Evaluation protocols.} Since \textsf{MM-GF} does not require parameter training, the notion of a conventional {\it training set} does not apply. Nevertheless, for consistency with widely adopted evaluation protocols, we randomly partition each user’s interaction history by using 80\% of interactions to construct the graph, 10\% for validation, and the remaining 10\% for testing. To assess the top-$K$ recommendation performance, we adopt the widely used metrics from prior studies \cite{DBLP:conf/sigir/0001DWLZ020, DBLP:conf/aaai/HeM16, DBLP:conf/mm/WeiWN0C20, DBLP:conf/mm/Zhang00WWW21, DBLP:conf/www/ZhouZLZMWYJ23, DBLP:conf/mm/ZhouS23, DBLP:conf/mm/Yu0LB23, DBLP:conf/cikm/ShenWZSZLL21, DBLP:conf/sigir/ParkSS24}, namely recall and normalized discounted cumulative gain (NDCG), where $K \in \left\{ {10,20} \right\} $. From RQ2 to RQ6 excecpt RQ4, we report the results only for NDCG@$20$ as similar trends are observed for other metrics. Note that \textsf{MM-GF} is fully deterministic; once the data split and hyperparameters are fixed, it produces identical predictions across runs. Therefore, we do not report statistics over multiple random seeds or conduct run-wise significance tests for \textsf{MM-GF}, and instead compare single-run results that are directly determined by the chosen configuration.

\noindent\textbf{Implementation details.} For a fair comparison, we implement the proposed \textsf{MM-GF} method and all benchmark methods using MMRec \cite{DBLP:conf/mmasia/Zhou23}, an open-sourced multimodal recommendation framework.\footnote{The toolbox is available at https://github.com/enoche/MMRec.} To calculate frequency bound of item--item similarity graphs, the minimum and maximum eigenvalues $(\lambda_{min}, \lambda_{max})$ are computed once per dataset and modality using the \texttt{torch.linalg.eigvalsh} routine. Unless otherwise stated, for \textsf{MM-GF}, the best-performing hyperparameters $(a_1, a_2, a_3, \beta, \gamma)$ in Eqs. \eqref{poly_filter} and \eqref{beta_gamma} are selected on the validation set from the following ranges: for the multimodal aggregation $P + \beta P_{\text{text}} + \gamma P_{\text{img}}$, we tune $\beta$ and $\gamma$ in $[0, 2]$ with a step size of $0.1$; for the polynomial filter, we search $a_1, a_2, a_3$ in $[-2, 2]$ with a step size of $0.1$. The finally chosen values are $(-0.1, 1.1, -0.7, 0.1, 0)$, $(1.5, 1.8, 0.7, 1.9, 0.5)$, and $(-0.1, 0.7, 1.9, 0.5, 0.2)$ for the Baby, Sports, and Clothing datasets, respectively. Additionally, we apply the proposed polynomial graph filter to the user--item interaction matrix, restricting the expansion to the third order (i.e., $K=3$ in Eq. \eqref{poly_filter}). For the textual and visual modalities, we adopt a linear LPF, specified as $\bar P_m^f = \bar P_m-\lambda_{min}^mI$, for each modality $m$. These design choices are primarily guided by empirical evidence. The parameter $k$ in Eq. \eqref{binary_similarity} was empirically set to 20 in our experiments. All experiments are carried out on a machine with Intel (R) 12-Core (TM) i7-9700K CPUs @ 3.60 GHz and an NVIDIA GeForce RTX A6000 GPU.

\subsection{Recommendation Accuracy (RQ1)}
\label{Accuracy}

Table \ref{main_table} summarizes the recommendation accuracy among \textsf{MM-GF} and seven recommendation competitors. Our observations are made as follows:
\begin{enumerate}[label=(\roman*)]
    \item Compared to state-of-the-art recommendation methods, \textsf{MM-GF} consistently achieves superior performance across all datasets and metrics. Notably, on the Baby dataset, \textsf{MM-GF} achieves up to a gain of 13.35\% in NDCG@$20$ over the second-best performer. This indicates that our nontrivial refinement for multimodal features and their effective fusion for GF enable us to achieve outstanding performance in MRSs.

    \item The GCN-based methods (LightGCN, GRCN, LATTICE, BM3, FREEDOM, and MGCN) generally outperform the non-GCN method (VBPR), highlighting the effectiveness of explicitly capturing high-order relations through the message passing mechanism in MRSs.

    \item Among the GCN-based methods, those utilizing multimodal information (GRCN, LATTICE, BM3, FREEDOM, and MGCN) exhibit better performance than their counterpart, i.e., the single-modal recommendation method (LightGCN). This underscores the importance of incorporating multimodal information for enhanced recommendation accuracy.
\end{enumerate}

\begin{table*}[t]\centering
\setlength
\tabcolsep{1.4pt}
\footnotesize
  \captionsetup{skip=3pt}
  \caption{Performance comparison among \textsf{MM-GF} and competitors. The best and second-best performers are highlighted in bold and underline, respectively. Here, R@$K$ and N@$K$ indicates Recall@$K$ and NDCG@$K$, respectively. We further conduct a user-level paired Wilcoxon signed-rank test, confirming that \textsf{MM-GF} significantly outperforms the strongest baseline (i.e., MGCN) on all evaluated datasets and metrics ($p \ll 0.01$).}
  \label{main_table}
  \resizebox{\textwidth}{!}{
  \begin{tabular}{c|cccc|cccc|cccc}
  \toprule
    \multicolumn{1}{c|}{}&\multicolumn{4}{c|}{Baby}&\multicolumn{4}{c|}{Sports}&\multicolumn{4}{c}{Clothing}\\
    \midrule
        Method &   R@10& N@10 &  R@20 & N@20 &  R@10& N@10 &  R@20 & N@20 &R@10& N@10 &  R@20 & N@20 \\
    \midrule
    LightGCN & 0.0479 & 0.0257 & 0.0754 & 0.0328 & 0.0569 & 0.0311 & 0.0864 & 0.0387 & 0.0340 & 0.0188& 0.0526  &0.0236\\
    VBPR & 0.0423 & 0.0223 & 0.0663 & 0.0284 & 0.0558 & 0.0307 & 0.0856 & 0.0384 & 0.0280 & 0.0159 & 0.0414  & 0.0193\\
    GRCN & 0.0539&0.0288&  0.0833 & 0.0363 & 0.0598& 0.0332& 0.0915 & 0.0414 & 0.0424& 0.0225& 0.0650  & 0.0283\\
    LATTICE & 0.0547& 0.0292 &0.0850 & 0.0370 & 0.0620&0.0335&  0.0953 & 0.0421 & 0.0492&0.0268& 0.0733  &0.0330\\
    BM3 & 0.0564&0.0301& 0.0883 & 0.0383 & 0.0656& 0.0355& 0.0980 & 0.0438 &0.0421&0.0228& 0.0625 & 0.0280 \\
    FREEDOM   & \underline{0.0627}&0.0330& \underline{0.0992}&  0.0424 &0.0717&0.0385& 0.1089 & 0.0481  & 0.0629& 0.0341& 0.0941  & 0.0420\\ 
    MGCN    & 0.0620&\underline{0.0339}& 0.0964 & \underline{0.0427} & \underline{0.0729}&\underline{0.0397}&  \underline{0.1106} & \underline{0.0496} & \underline{0.0641}&\underline{0.0347} & \underline{0.0945} & \underline{0.0428}\\
    \midrule
    \textbf{MM-GF}& \textbf{0.0757} & \textbf{0.0432} & \textbf{0.1109}& \textbf{0.0522} & \textbf{0.0849} & \textbf{0.0495} & \textbf{0.1226} & \textbf{0.0593} & \textbf{0.0761} & \textbf{0.0424} & \textbf{0.1108} & \textbf{0.0513}\\
    \bottomrule
  \end{tabular}
  }
\end{table*}

\subsection{Runtime and Scalability (RQ2)}
Table \ref{runtime_table} summarizes the runtime of \textsf{MM-GF} and GCN-based competitors that perform well (GRCN, LATTICE, BM3, and MGCN) on the three datasets. For the GCN-based methods, runtime refers to the training time, whereas, for the GF-based method (\textsf{MM-GF}), it indicates the processing time, as in \cite{DBLP:conf/cikm/ShenWZSZLL21, DBLP:conf/sigir/ParkSS24}. We observe that, on the Baby dataset, \textsf{MM-GF} performs approximately \underline{$\times100.4$ faster} than MGCN, which is the best GCN-based multimodal recommendation method, while exhibiting even higher recommendation accuracy. This is because \textsf{MM-GF} operates solely on straightforward matrix calculations without a costly training process. A similar tendency is observed on other two datasets. In consequence, the proposed \textsf{MM-GF} method is advantageous in terms of runtime as well as recommendation accuracy.

\begin{figure}[t]
    \centering
    \includegraphics[width=0.6\columnwidth]{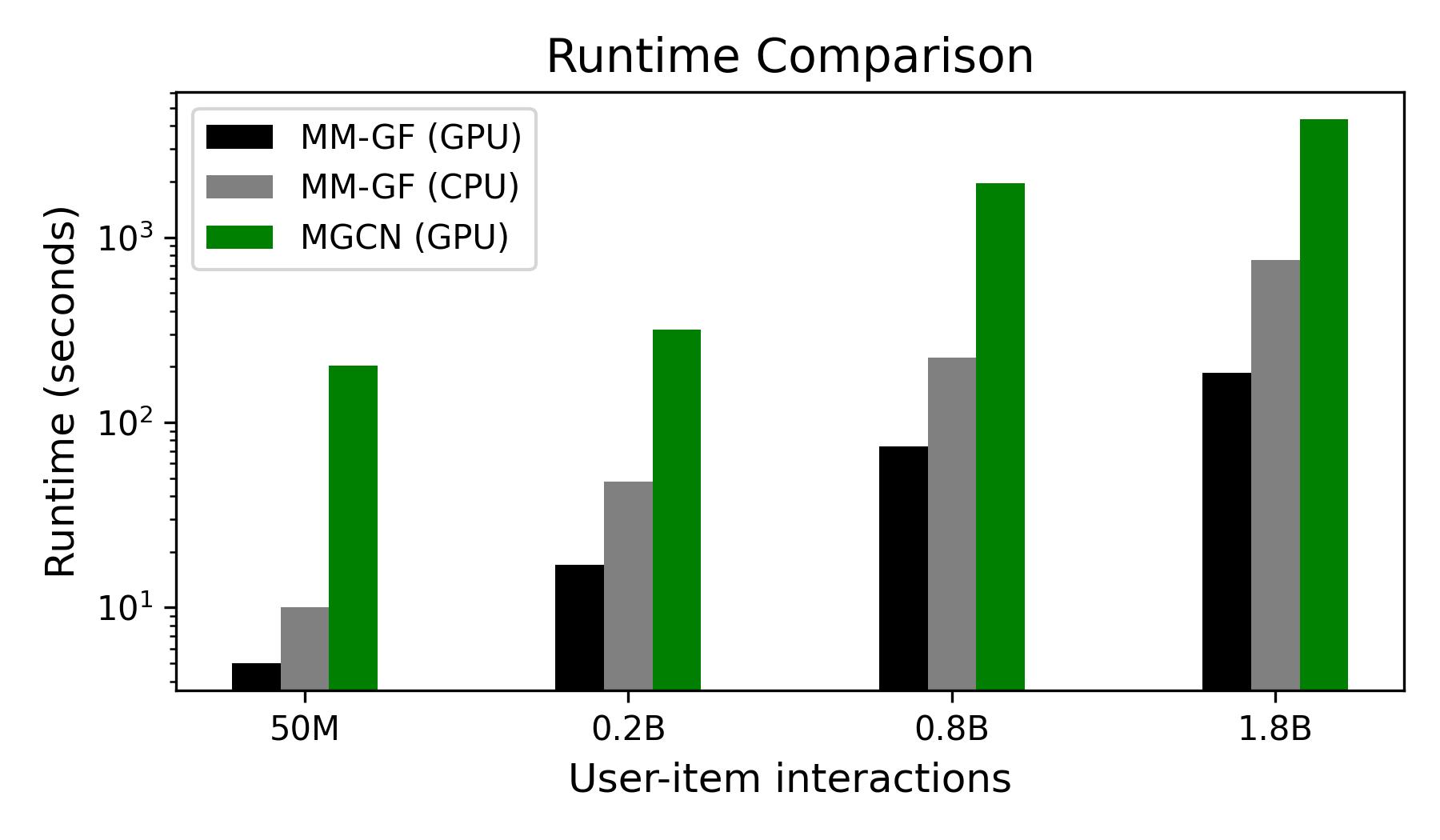}
    \vspace{0mm}
    \caption{Log-scaled runtime comparison of \textsf{MM-GF} (with GPU and CPU) and MGCN (GPU) using various scaled synthetic datasets.}
    \vspace{-5mm}
    \label{runtime2}
\end{figure}

Furthermore, we compare the scalability of \textsf{MM-GF} and MGCN, which is the fastest and best-performing one out of GCN-based methods. We present a runtime comparison on different devices (i.e., CPU and GPU), using various scaled datasets. To this end, we generate four synthetic datasets whose sparsity is identically set to $99.99\%$, similarly as in three real-world benchmark datasets, i.e., Baby, Sports, and Clothing. More specifically, the numbers of (users, items, interactions) are set to \{(10k, 5k, 5k), (20k, 10k, 20k), (40k, 20k, 80k), (60k, 30k, 180k)\}. Additionally, to match the dimensionality of the feature embeddings, we set the dimensionality of textual and visual features to 384 and 4,096, respectively. As shown in Fig. \ref{runtime2}, \textsf{MM-GF} has significantly shorter runtime than that of MGCN for all datasets and device configurations. Interestingly, running \textsf{MM-GF} with CPU is even faster than the case of MGCN with GPU. We also observe that \textsf{MM-GF} with GPU takes at most few minutes, while MGCN takes a minimum of several minutes and a maximum of several hours.

\begin{table}[t]
\footnotesize
\centering
\captionsetup{skip=2pt}
\caption{Runtime comparison among \textsf{MM-GF} and representative GCN-based competitors. The best performer is highlighted in bold. N@20 refers to NDCG@$20$.}
\label{runtime_table}
\begin{tabularx}{\linewidth}{l|XX|XX|XX}
\toprule
\multicolumn{1}{c|}{}&\multicolumn{2}{c|}{Baby}&\multicolumn{2}{c|}{Sports}&\multicolumn{2}{c}{Clothing}\\
\midrule
Method & N@20 & Time & N@20 & Time & N@20 & Time \\
\midrule
GRCN&0.0363&3h8m&0.0414&5h34m&0.0283&6h14m\\
LATTICE&0.0370&2h19m&0.0421&17h44m&0.0330&8h56m\\
BM3&0.0383&1h27m&0.0438&2h55m&0.0280&2h39m\\
MGCN&0.0427&13m33s&0.0496&58m4s&0.0428&53m11s\\
\midrule
\textsf{MM-GF}&\textbf{0.0522}&\textbf{8.1s}&\textbf{0.0593}&\textbf{1m16s}&\textbf{0.0513}&\textbf{2m14s}\\
\bottomrule
\end{tabularx}

\end{table}

\begin{table}[t]
\footnotesize
\centering
\captionsetup{skip=2pt}
\caption{Performance comparison among \textsf{MM-GF} and its four variants in terms of NDCG@$20$. The best performer is highlighted in bold.} 
\label{table:ablation}
\begin{tabularx}{\linewidth}{l|Y|Y|Y|Y|Y|Y}
\toprule
Dataset & \smash{\textsf{MM-GF}} & \smash{\textsf{MM-GF-t}} & \smash{\textsf{MM-GF-tv}} & \smash{\textsf{MM-GF-pf}} & \smash{\textsf{MM-GF-f}} & \smash{\textsf{MM-GF-h}}\\ 
\midrule
Baby & \textbf{0.0522} & 0.0448 $(\downarrow 14.2\%)$ & 0.0448 $(\downarrow 14.2\%)$ & 0.0484 $(\downarrow 7.3\%)$ & 0.0420 $(\downarrow 19.5\%)$ & 0.0417 $(\downarrow 20.1\%)$ \\
Sports & \textbf{0.0593} & 0.0493 $(\downarrow 16.9\%)$ & 0.0485 $(\downarrow 18.2\%)$ & 0.0562 $(\downarrow 5.2\%)$ & 0.0486 $(\downarrow 18.0\%)$ & 0.0429 $(\downarrow 27.7\%)$ \\
Clothing & \textbf{0.0513} & 0.0377 $(\downarrow 26.5\%)$ & 0.0348 $(\downarrow 32.2\%)$  & 0.0445 $(\downarrow 13.3\%)$ & 0.0308 $(\downarrow 40.0\%)$ & 0.0394 $(\downarrow 23.2\%)$ \\
\bottomrule
\end{tabularx}
\vspace{-4mm}
\end{table}

\definecolor{linecol1}{rgb}{1.0, 0.5, 0.0}
\definecolor{linecol2}{rgb}{0.1, 0.6, 0.0}
\definecolor{linecol3}{rgb}{0.2, 0.4, 0.8}
\definecolor{linecol4}{rgb}{0.2, 0.1, 0.7}
\definecolor{linecol5}{rgb}{0.8, 0.1, 0.3}

\subsection{Ablation Study (RQ3)} \label{sec5.3} 

We perform an ablation study to assess the contribution of each component in \textsf{MM-GF}. The performance comparison among \textsf{MM-GF} and its four variants is summarized in Table \ref{table:ablation}.
\begin{itemize}
    \item \textsf{MM-GF} : preserves all components with no removal;
    \item \textsf{MM-GF-t} : excludes the textual feature (i.e., $\beta = 0$);
    \item \textsf{MM-GF-tv} : excludes the textual and visual features (i.e., $ \beta = \gamma = 0$);
    \item \textsf{MM-GF-pf} : excludes the polynomial filter for user--item interaction data (i.e., adopts a linear LPF);
    \item \textsf{MM-GF-f} : rather uses the fixed coefficients $a_k$ in Eq. \eqref{poly_filter} as in \cite{DBLP:conf/sigir/ParkSS24, DBLP:conf/www/ParkYS25, DBLP:conf/www/KimCPS25};
    \item \textsf{MM-GF-h} : constructs the model without spectral bound adjustment in Eq. \eqref{poly_filter}, i.e., directly applies the polynomial filter to the original spectrum as in \cite{DBLP:conf/sigir/ParkSS24, DBLP:conf/www/ParkYS25, DBLP:conf/www/KimCPS25}.
\end{itemize}

Our observations are made as follows:

\begin{enumerate}[label=(\roman*)]
    \item  The presence of multimodal information significantly contributes to performance improvement for all cases. In particular, the textual modality has a substantial impact on recommendation accuracy.

    \item As evidenced by the results of both \textsf{MM-GF-t} and \textsf{MM-GF-tv}, the gain of further leveraging the visual modality over \textsf{MM-GF-t} is indeed marginal; such a pattern was also comprehensively discussed in the earlier study in \cite{zhou2023comprehensive}.

    \item The performance degradation in \textsf{MM-GF-pf} demonstrates that our modality-specific polynomial filter design contributes to the improvement of performance. Moreover, handling the bound of frequencies is essential for accurate predictions, as \textsf{MM-GF-f} and \textsf{MM-GF-h} lead to a considerable drop in performance.

\end{enumerate}

\begin{table*}[t]\centering
\setlength
\tabcolsep{1.4pt}
\footnotesize
  \captionsetup{skip=3pt}
  \caption{Performance comparison among \textsf{MM-GF} and competitors in cold-start setting. The best and second-best performers are highlighted in bold and underline, respectively. Here, R@$K$ and N@$K$ indicates Recall@$K$ and NDCG@$K$, respectively.}
  \label{cold_table}
  \resizebox{\textwidth}{!}{
  \begin{tabular}{c|cccc|cccc|cccc}
  \toprule
    \multicolumn{1}{c|}{}&\multicolumn{4}{c|}{Baby}&\multicolumn{4}{c|}{Sports}&\multicolumn{4}{c}{Clothing}\\
    \midrule
        Method &   R@10& N@10 &  R@20 & N@20 &  R@10& N@10 &  R@20 & N@20 &R@10& N@10 &  R@20 & N@20 \\
    \midrule
    LightGCN & 0.0400 & 0.0216 & 0.0646 & 0.0278 & 0.0434 & 0.0236 & 0.0642 & 0.0288 & 0.0256 & 0.0143 & 0.0380 & 0.0174\\
    VBPR & 0.0199 & 0.0111 & 0.0320 & 0.0138 & 0.0205 & 0.0113 & 0.0302 & 0.0138 & 0.0196 & 0.0102 & 0.0308  & 0.0130\\
    GRCN & 0.0301&0.0158&  0.0480 & 0.0203 & 0.0342& 0.0180& 0.0543 & 0.0230 & 0.0306& 0.0161& 0.0474  & 0.0203\\
    LATTICE & 0.0424& 0.0234 &0.0656 & 0.0292 & 0.0525&0.0285&  0.0771 & 0.0347 & 0.0481&0.0267& 0.0679  &0.0317\\
    BM3 & 0.0323&0.0165& 0.0468 & 0.0202 & 0.0425& 0.0237& 0.0603 & 0.0281 &0.0261& 0.0146& 0.0363 & 0.0172 \\
    FREEDOM   & 0.0463&0.0258& \underline{0.0711}& \underline{0.0321} & 0.0604 & 0.0328 & \underline{0.0909} & 0.0404  & \underline{0.0591} &\underline{0.0323}& \underline{0.0883} & \underline{0.0396}\\ 
    MGCN    & \underline{0.0482}&\underline{0.0259}& 0.0700 & 0.0313 & \underline{0.0619} & \underline{0.0351} & 0.0900 & \underline{0.0422} & 0.0540 & 0.0302& 0.0771 & 0.0361\\
    \midrule
    \textbf{MM-GF}& \textbf{0.0629} & \textbf{0.0349} & \textbf{0.0928}& \textbf{0.0424} & \textbf{0.0757} & \textbf{0.0433} & \textbf{0.1056} & \textbf{0.0509} & \textbf{0.0673} & \textbf{0.0362} & \textbf{0.0975} & \textbf{0.0439}\\
    \bottomrule
    \end{tabular}
    }
\end{table*}


\subsection{Analysis in Cold-Start Settings (RQ4)}

While high sparsity in graphs leads to technical challenges for model training, leveraging additional features in MRSs can enhance the recommendation accuracy in sparse conditions, as demonstrated by cold-start experiments \cite{zhou2023comprehensive}. In this study, we regard the cold-start users as those who have rated equal to or fewer than 5 items. Hence, we only use such cold-start users on three benchmark datasets for model inference. The performance comparison among \textsf{MM-GF} and seven state-of-art multimodal recommendation competitors in cold-start settings is summarized in Table \ref{cold_table}.\footnote{Since the datasets in cold-start settings differ from the original ones, we determine the optimal hyperparameters for each dataset under these settings.} Our observations are made as follows:

\begin{enumerate}[label=(\roman*)]
    \item  Compared to the GCN-based methods (LightGCN, GRCN, LATTICE, BM3, FREEDOM, and MGCN), \textsf{MM-GF} still consistently achieves superior performance across all datasets and metrics (except for the Recall@$20$ on Clothing). Notably, on the Baby dataset, \textsf{MM-GF} achieves up to a gain of $35.46\%$ in NDCG@$20$ over the best competitor.
    
    \item  In comparison with Table \ref{main_table}, the previous multimodal recommendation methods such as VBPR, GRCN, and BM3 have a large performance decrease, showing vulnerability to cold-start settings. On the other hand, our \textsf{MM-GF} method has a marginal performance decrease in cold-start experiments. This finding demonstrates the robustness of our \textsf{MM-GF} method to cold-start settings.

    \item From a modeling perspective, this robustness can be attributed to the graph-filtering nature of \textsf{MM-GF}. Since trainable GNN-based recommenders \cite{DBLP:conf/aaai/HeM16, DBLP:conf/mm/WeiWN0C20, DBLP:conf/mm/Zhang00WWW21, DBLP:conf/www/ZhouZLZMWYJ23, DBLP:conf/mm/ZhouS23, DBLP:conf/mm/Yu0LB23} rely on parameter updates that are directly driven by historical interactions, items with very few edges receive weak or unstable updates and tend to be poorly represented in cold-start regimes. In contrast, \textsf{MM-GF} constructs modality-specific item--item similarity graphs and then applies a polynomial graph filter to these graphs; even for cold-start items with sparse interactions, their representations are smoothed toward those of spectrally similar and well-connected neighbors. Consequently, the effective information available to such items is enriched through GF, which explains why \textsf{MM-GF} maintains relatively stable performance under sparse (cold-start) conditions.
\end{enumerate}

\subsection{Robustness to Noisy Multimodal Features (RQ5)}
\label{Robustness noise}

We analyze the sensitivity of noise that often occurs in real-world scenarios \cite{zhou2023comprehensive}. In MRSs, multimodal features are vulnerable to noise due to various factors such as embedding inaccuracies or inconsistencies in data collection processes. To validate the robustness of \textsf{MM-GF} to such noisy multimodal features, we characterize the noise $n$ as $\tilde{X}_m^n=\tilde{X}_m + n$, where $n$ follows $\mathcal{N}(0, \sigma_m^2)$ for the standard deviation $\sigma_m$. We define six different levels of noise according to different levels of the standard deviation: level 0 corresponds to the case where there is no noise, which represents the original datasets; level 1 corresponds to the noise equivalent to $10\%$ of the standard deviation of each feature embedding; and as noise gets gradually added, level 5 corresponds to the noise that is twice the standard deviation of each feature embedding.

\begin{figure}[t]
    \centering
    \includegraphics[width=0.8\columnwidth]{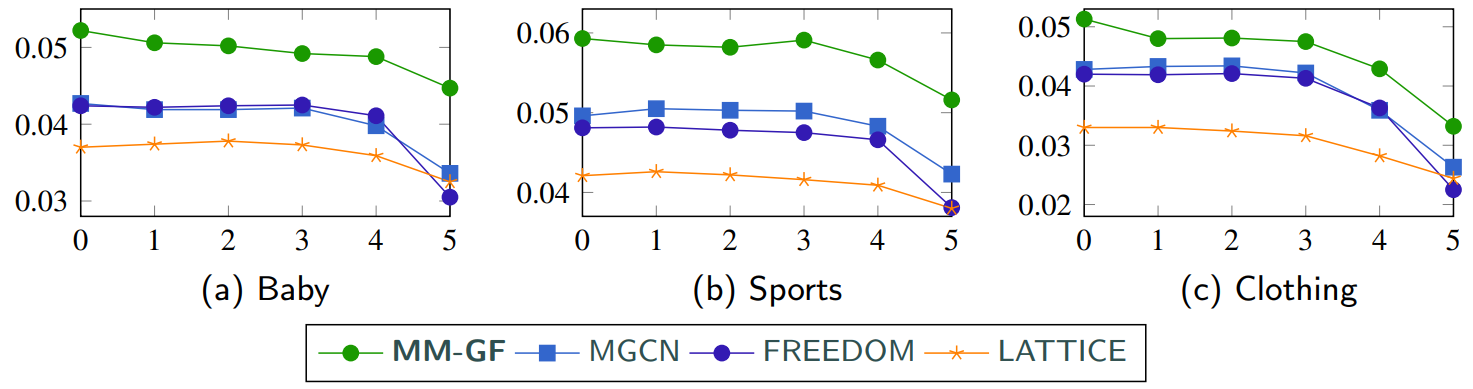}
    \vspace{0mm}
    \caption{Performance comparison according to different degrees of noise. Here, the horizontal axis indicates the noise level $x \in \{0,1,2,...,5\}$, which is specified in Section \ref{Robustness noise}. The vertical axis means NDCG@$20$.}
    \vspace{-5mm}
    \label{noise}
\end{figure}

Figure \ref{noise} clearly shows that as the multimodal information incorporates more noise, all the competitors exhibit a decreasing trend in NDCG@$20$. In particular, MGCN \cite{DBLP:conf/mm/Yu0LB23} purifies the modality features to prevent noise contamination, which is a method directly designed for noise removal. In contrast, \textsf{MM-GF} consistently reveals the best performance in MRSs through only matrix computations without any noise-purification design. In other words, \textsf{MM-GF} exhibits robustness to noise with a simple yet effective multimodal feature process.

From a spectral viewpoint, this robustness can be explained by the design of \textsf{MM-GF}. Additive Gaussian noise on multimodal feature embeddings mainly manifests as high-frequency components \cite{DBLP:journals/spm/ShumanNFOV13, DBLP:conf/icml/YuQ20}, whereas the proposed polynomial graph filter behaves as a controlled low-pass operator on the bounded spectrum. By attenuating high-frequency fluctuations while preserving the dominant low-frequency structure, \textsf{MM-GF} effectively suppresses noise in multimodal features without any explicit denoising module. Moreover, since our method is training-free and does not learn a large number of parameters from noisy input, it is less prone to overfitting to noise than heavily parameterized GNN-based models, which further contributes to its stable performance under noisy conditions.

\begin{figure}[t]
    \centering
    \begin{subfigure}[b]{0.32\linewidth}
        \centering
        \includegraphics[width=\linewidth]{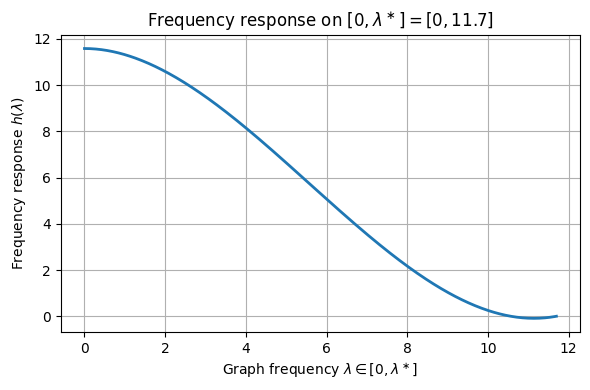}
        \caption{Baby}
    \end{subfigure}
    \begin{subfigure}[b]{0.32\linewidth}
        \centering
        \includegraphics[width=\linewidth]{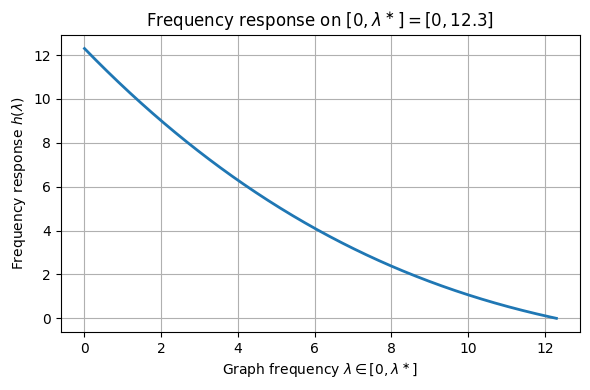}
        \caption{Sports}
    \end{subfigure}
    \begin{subfigure}[b]{0.32\linewidth}
        \centering
        \includegraphics[width=\linewidth]{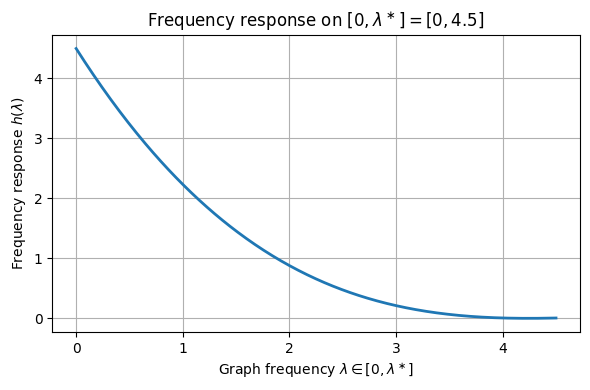}
        \caption{Clothing}
    \end{subfigure}

    \captionsetup{skip=-2pt}
    \vspace{4mm}
    \caption{Frequency response functions tuned on three datasets with respect to the graph frequency $\lambda$.}
    \label{frequency response}
\end{figure}

\begin{figure}[t]
    \centering
    \begin{subfigure}{0.48\textwidth}
        \centering
        \includegraphics[width=\linewidth]{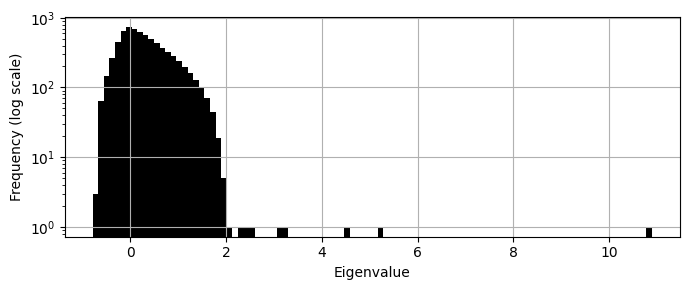}
        \caption{Unbounded 1st-order polynomial LPF}
        \label{fig:unbounded range 1st}
    \end{subfigure}
    \hfill
    \begin{subfigure}{0.48\textwidth}
        \centering
        \includegraphics[width=\linewidth]{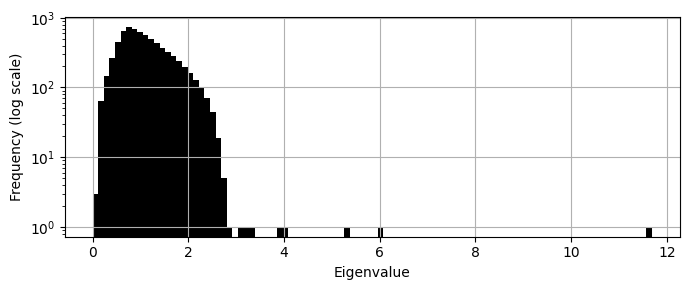}
        \caption{Bounded 1st-order polynomial LPF}
        \label{fig:bounded range 1st}
    \end{subfigure}
    \begin{subfigure}{0.48\textwidth}
        \centering
        \includegraphics[width=\linewidth]{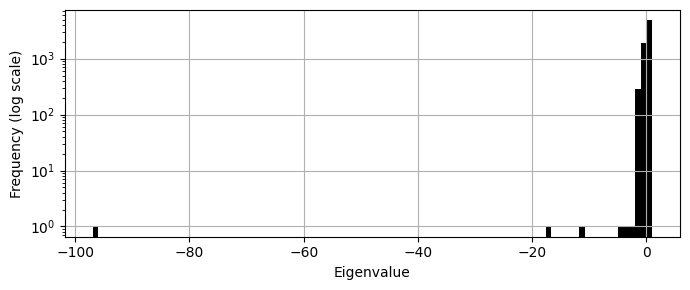}
        \caption{Unbounded 2nd-order polynomial LPF}
        \label{fig:unbounded range 2nd}
    \end{subfigure}
    \hfill
    \begin{subfigure}{0.48\textwidth}
        \centering
        \includegraphics[width=\linewidth]{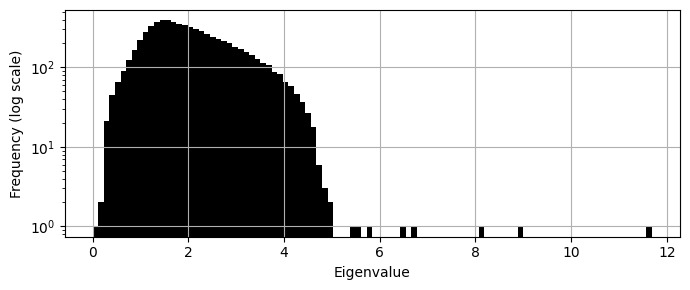}
        \caption{Bounded 2nd-order polynomial LPF}
        \label{fig:bounded range 2nd}
    \end{subfigure}
    \begin{subfigure}{0.48\textwidth}
        \centering
        \includegraphics[width=\linewidth]{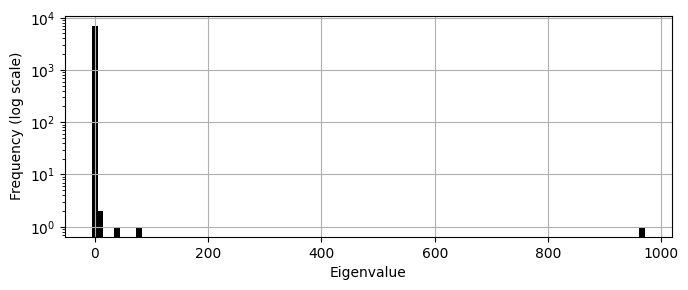}
        \caption{Unbounded 3rd-order polynomial LPF}
        \label{fig:unbounded range 3rd}
    \end{subfigure}
    \hfill
    \begin{subfigure}{0.48\textwidth}
        \centering
        \includegraphics[width=\linewidth]{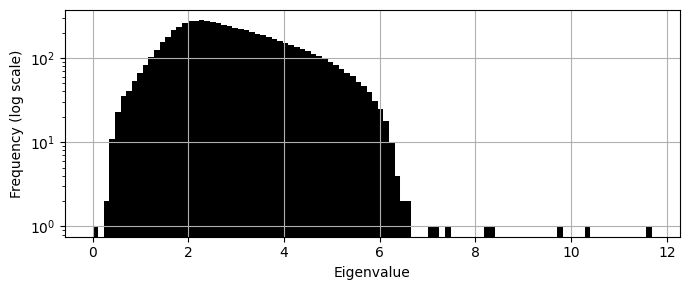}
        \caption{Bounded 3rd-order polynomial LPF}
        \label{fig:bounded range 3rd}
    \end{subfigure}
    \caption{Eigenvalue distributions under different polynomial LPFs, with left figures showing unbounded spectra without our spectral regularization and right ones illustrating well-bounded spectra after applying the proposed normalization for 1st-, 2nd-, and 3rd-order polynomial filters, respectively.}
    \vspace{-4mm}
\end{figure}

\subsection{Spectral Analysis (RQ6)}
\label{Spectral analysis}

We examine how the polynomial graph filter of \textsf{MM-GF} operates under different spectral conditions across datasets. Figure \ref{frequency response} illustrates the frequency response functions derived from the user--item interaction graphs of three datasets with respect to the graph frequency $\lambda$. Owing to their distinct eigenvalue distributions, each dataset induces a different spectral range $[0,\lambda^*]$, which leads the filter to adopt dataset-specific attenuation patterns. This demonstrates that the proposed polynomial filter adapts its spectral emphasis to the underlying graph structure rather than imposing a fixed response, thereby highlighting that the adaptable polynomial coefficients of \textsf{MM-GF} play a central role in shaping its behavior under different spectral conditions.

Additionally, we analyze the eigenvalue convergence induced by the proposed spectral regularization. Figure 8 compares the eigenvalue distributions before and after applying our spectral normalization in Eq. \eqref{poly_filter}. When the polynomial filters are applied \textit{without} our regularization (see Figures \ref{fig:unbounded range 1st}, \ref{fig:unbounded range 2nd} and \ref{fig:unbounded range 3rd}), the resulting spectra easily escape the desirable range $[0, \lambda^{\ast}]$. This issue becomes dramatically evident for higher-order filters. For example, the 3rd-order polynomial LPF $h(\lambda) = 1 - \lambda^3$ produces eigenvalues approaching $1,000$, which makes the filter unstable and undermines any meaningful interpretation of the spectral response. Such divergence clearly indicates that unbounded polynomial filters cannot operate reliably on real-world user--item graphs, and the spectral domain quickly becomes distorted.

In contrast, Figures \ref{fig:bounded range 1st}, \ref{fig:bounded range 2nd}, and \ref{fig:bounded range 3rd} show the spectra obtained \textit{after} applying our spectral regularization, which rescales the graph operator so that all polynomial filters (1st-, 2nd-, and 3rd-order polynomials) produce eigenvalues that remain strictly within a compact interval. This boundedness demonstrates the convergence stability of the proposed method; that is, regardless of polynomial orders, the filtered operator does not amplify spectral components beyond the specific range. Consequently, the proposed polynomial graph filters preserve their intended low-pass characteristics and remain interpretable in the spectral domain.

\begin{figure}[t]
    \centering
    \includegraphics[width=0.9\columnwidth]{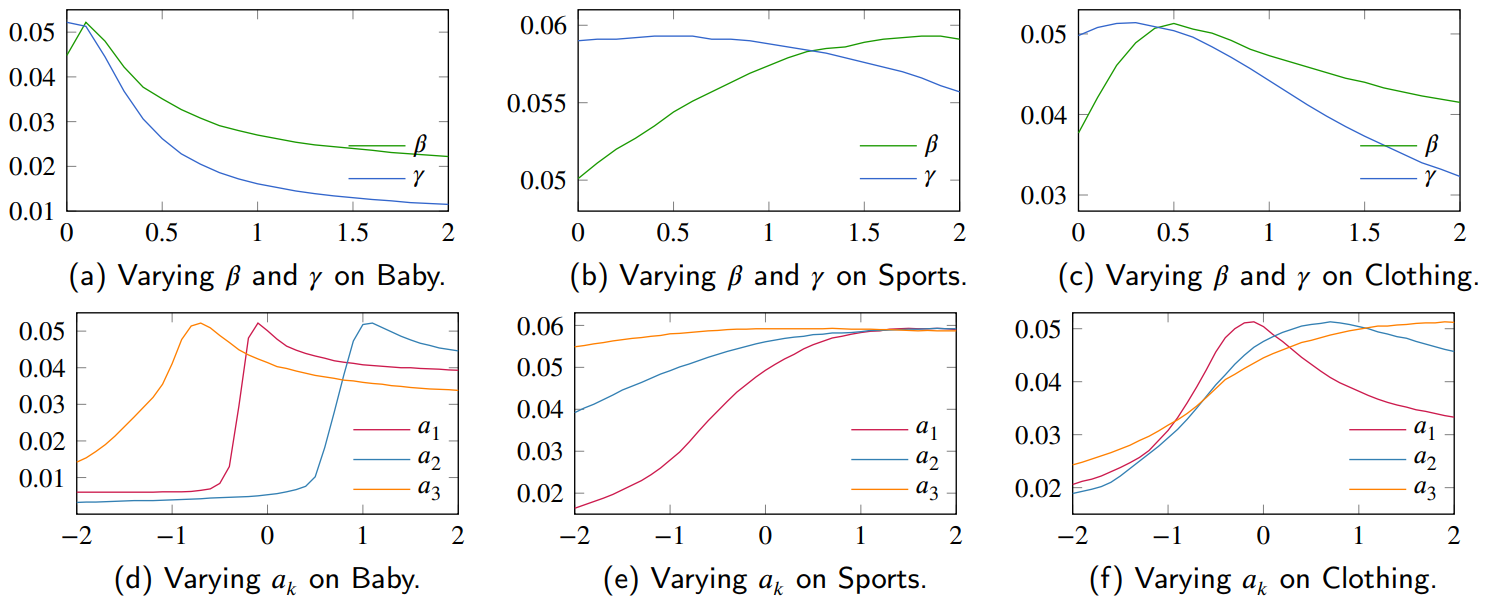}
    \vspace{0mm}
    \caption{The effect of various hyperparameters for three benchmark datasets, where the horizontal and vertical axes indicate the value of each hyperparameter and the performance in NDCG@$20$, respectively. Each plot varies a single hyperparameter while keeping the others fixed.}
    \vspace{-5mm}
    \label{hyper_plot}
\end{figure}

\subsection{Hyperparameter Sensitivity (RQ7)}
\label{hyperparameter sensitivity}

We analyze the sensitivity of \textsf{MM-GF} on the performance in NDCG@$20$ to variations in the key hyperparameters $\beta$,$\gamma$, and $a_k$ in Figure \ref{hyper_plot}.\footnote{We note that the differing scales of $\beta$ and $\gamma$ in the optimization process stem from the heterogeneous nature of multimodal features, while the variations in $a_k$ arise from the flexibility of the polynomial filter design.} Our observations are made as follows:

\begin{enumerate}[label=(\roman*)]
    \item Setting $\beta$ and $\gamma$ to positive values generally leads to higher accuracies than the case of $\beta=0$ and $\gamma=0$, thus validating the effectiveness of leveraging both textual and visual features for GF.

    \item As shown in Figure \ref{hyper_plot}a, increasing the weight $\gamma$ of the visual modality leads to a consistent performance drop, suggesting that visual features may introduce noise rather than benefit recommendation accuracy. This observation aligns with prior findings \cite{zhou2023comprehensive}, which report that incorporating all modalities does not universally improve performance and that the usefulness of visual features varies depending on dataset characteristics and feature quality.

    \item While the optimal values of hyperparameters differ across datasets and necessitate a broad search range, \textsf{MM-GF} is inherently less sensitive to hyperparameter choices and allows for rapid tuning. This is in contrast to training-based methods \cite{DBLP:conf/mm/WeiWN0C20,DBLP:conf/mm/Zhang00WWW21, DBLP:conf/mm/Yu0LB23, DBLP:conf/mm/ZhouS23}, which often demand costly and time-consuming optimization procedures. Our model’s simplicity enables efficient deployment even with minimal tuning effort.
\end{enumerate}

\begin{table}[t]
\centering
\caption{The performance according to the polynomial order $K$ for the user--item interaction graph with respect to NDCG@20. The best performer is highlighted in bold.}
\label{K_diff}
\begin{tabular}{l|ccc}
\toprule
Dataset  & $K = 1$  & $K = 2$  & $K = 3$ \\
\midrule
Baby     & 0.0465   & 0.0499   & \textbf{0.0522} \\
Sports   & 0.0547   & 0.0581   & \textbf{0.0593} \\
Clothing & 0.0448   & 0.0467   & \textbf{0.0513} \\
\bottomrule
\end{tabular}
\end{table}

Furthermore, to validate the choice of the polynomial order, we conduct an additional experiment by varying the orders for the similarity graph constructed from user--item interactions. In case of multimodal information, we set the polynomial order to 1, as our experiments indicate that further increasing the orders not only introduces additional hyperparameters but also leads to performance degradation. We vary the polynomial order as $K \in \{1,2,3\}$ while keeping all other settings fixed. As summarized in Table \ref{K_diff}, increasing $K$ from 1 to 3 yields consistent improvements in NDCG@20 across all three datasets. We do not extend the analysis to $K > 3$, since higher order introduce additional polynomial coefficients and require more matrix multiplications, resulting in substantially higher computational cost and, in practice, undermine the efficiency benefits expected from a training-free framework.

\section{Conclusions and Outlook}
\label{section 6}
In this paper, we proposed \textsf{MM-GF}, the first training-free multimodal recommendation method grounded in the concept of GF, aiming for both efficiency and accuracy. \textsf{MM-GF} effectively adjusted the spectral bounds of frequencies, alongside a flexible and spectrally-aware GF strategy. By integrating modality-specific similarity graphs with a tunable polynomial graph filter, our \textsf{MM-GF} method was shown to enable precise frequency control and robust fusion of heterogeneous information. Extensive experimental evaluations demonstrated not only the remarkably fast runtime of \textsf{MM-GF} but also the superior recommendation accuracy of \textsf{MM-GF} in diverse challenging scenarios, including cold-start conditions and resilience to noisy features. Potential avenues of future research include adaptive mechanisms to make the filtering and modality fusion more data-dependent. Future research in this area in the inclusion of an extension to streaming recommendation, where the training-free nature of \textsf{MM-GF} allows frequent refresh of recommendations under rapidly changing interactions and contents.

Nevertheless, \textsf{MM-GF} has several challenges and limitations. First, since we directly use pre-trained textual and visual embeddings without additional modality-specific pre- or post-processing, the contribution of multimodal information can be limited to some extent when these features are noisy or weak. Second, the static design of \textsf{MM-GF}—where filter coefficients and modality weights are fixed after validation—ensures efficiency and simplicity, but requires manual retuning under shifts in data distribution or modality balance. We leave addressing these issues---for example, by incorporating lightweight adaptive mechanisms or more sophisticated multimodal processing while preserving the training-free spirit---as important directions for future work.

\section*{Acknowledgments}
This work was supported in part by the National Research Foundation of Korea (NRF) grant funded by the Korea government (MSIT) (No. RS-2021-NR059723, No. RS-2023-00220762) and by SMEs Technology Innovation Development Program through the Technology Innovation and Promotion Agency (TIPA), funded by Ministry of SMEs and Startups (RS-2024-00511332).

\bibliographystyle{cas-model2-names}

\bibliography{cas-sc-template}



\end{document}